\documentclass[prd,aps,a4paper,floatfix,amsmath,amssymb,twocolumn,nofootinbib]{revtex4-1}
\usepackage{amssymb,amsmath,xcolor,graphicx}
\usepackage[margin=2cm]{geometry}
\usepackage[pdftex,breaklinks,colorlinks,
linkcolor=blue,
citecolor=teal,
anchorcolor=red,
urlcolor=cyan]{hyperref}

\newcommand{\R}{{\mathbb R}}
\newcommand{\Z}{{\mathbb Z}}
\newcommand{\x}{{\bf x}}
\newcommand{\y}{{\bf y}}
\newcommand{\z}{{\bf z}}
\DeclareMathOperator{\tr}{tr}

\begin{document}

\title{The two-body problem in 2+1 spacetime dimensions with negative
  cosmological constant: two point particles}

\author{Carsten Gundlach} 
\affiliation{School of Mathematical
  Sciences, University of Southampton, Southampton SO17 1NY, United
  Kingdom}

\date{5 July 2024}


\begin{abstract}

We work towards the general solution of the two-body problem in
2+1-dimensional general relativity with a negative cosmological
constant. The BTZ solutions corresponding to black holes, point
particles and overspinning particles can be considered either as
objects in their own right, or as the exterior solution of compact
objects with a given mass $M$ and spin $J$, such as rotating fluid
stars. We compare and contrast the metric approach to the
group-theoretical one of characterising the BTZ solutions as
identifications of 2+1-dimensional anti-de Sitter spacetime under an
isometry. We then move on to the two-body problem. In this paper, we
restrict the two objects to the point particle range $|J|-1\le
M<-|J|$, or their massless equivalents, obtained by an infinite
boost. (Both anti-de Sitter space and massless particles have $M=-1$,
$J=0$). We derive analytic expressions for the total mass
$M_\text{tot}$ and spin $J_\text{tot}$ of the system in terms of the
six gauge-invariant parameters of the two-particle system: the rest
mass and spin of each object, and the impact parameter and energy of
the orbit. Based on work of Holst and Matschull on the case of two
massless, nonspinning particles, we conjecture that the black hole
formation threshold is $M_\text{tot}=|J_\text{tot}|$. The threshold
solutions are then extremal black holes. We determine when the global
geometry is a black hole, an eternal binary system, or a closed
universe.

\end{abstract}


\maketitle

\tableofcontents


\section{Introduction}
\label{section:introductionI}


General relativity in 2+1 spacetime dimensions appears dynamically
trivial because in 2+1 dimensions the Weyl tensor is identically
zero. This means that the full Riemann tensor is determined by the
Ricci tensor, and so by the stress-energy tensor of the matter. Hence
there are no gravitational waves, and the vacuum solution is locally
unique: Minkowski in the absence of a cosmological constant $\Lambda$,
de Sitter for $\Lambda>0$ and anti-de Sitter (from now on, adS3) for
$\Lambda<0$.

However, it was noted by Deser, Jackiw and t'Hooft in 1984
\cite{DeserJackiwTHooft84} that even the general vacuum solution with
$\Lambda=0$ is non-trivial if one allows for singularities
representing point particles, which may be spinning. The global
solution is then obtained by identifying Minkowski spacetime with
itself under a non-trivial isometry for each particle. In the simplest
case, this is a rotation by $2\pi-2\nu$, where the resulting
spacetime can be thought of as Minkowski with a wedge of opening angle
$2\nu$ removed and the two faces of the wedge identified. The
dynamics of two or more interacting particles can then be dealt with
in closed form using the algebra of such isometries.

In 1992, Ba\~nados, Teitelboim and Zanelli \cite{BTZ92} (from now on,
BTZ) noticed that 2+1 dimensional vacuum Einstein gravity with
$\Lambda<0$ admits rotating black hole solutions parameterised by a
mass $M$ and spin $J$ that share many features with the family of Kerr
solutions in 3+1 dimensions. These can be found easily by solving an
axistationary ansatz for the metric, but their existence had been
overlooked because the metric has to be locally adS3. In fact, these
metrics can be derived as non-trivial identifications of adS3 with
itself \cite{BHTZ93} under the action of an isometry.

A key difference to black holes in 3+1 dimensions is the existence of
a mass gap: while adS3 is given by the BTZ solution with parameters
$M=-1$ and $J=0$, only the BTZ solutions with $M\ge0$ and $|J|\le M$
represent black holes. (These parameters are defined below). Solutions
with all other real values of $M$ and $J$ have naked singularities and
represent point particles ($|J|<-M$), similar to those for $\Lambda=0$
described in \cite{DeserJackiwTHooft84}, and overspinning particles
($|J|>|M|$). 

Combining these two ideas suggests a research programme of completely
solving the two-body problem in 2+1-dimensional gravity with
$\Lambda<0$.  A key insight is that the BTZ solutions are not only
building blocks in their own right, but are also the exterior
solutions of any compact object. By contrast, in 3+1 dimensions, the
vacuum exterior spacetime of any spherical compact object is
Schwarzschild, but the exterior of a non-spherical or rotating compact
object is not in general Kerr, even the if the object is
itself axistationary.

Because any vacuum spacetime with $\Lambda<0$ is locally isometric to
adS3, the exterior spacetime of any compact object, like the vacuum
BTZ solutions, must be an identification of adS3 with itself under an
element of its isometry group. The basic idea is the following. Cover
up the compact object with a world tube, remove the interior of the
world tube, and make the resulting spacetime simply connected by
making a suitable cut from the world tube to infinity. On this simply
connected domain, the spacetime must now locally be adS3. To undo the
cutting-open, we need to identify the two sides of the cut, and
regularity then requires that the identification is via an isometry of
adS3. It was shown in \cite{BHTZ93} that the gauge-invariant content
of such an identification is a mass $M$ and spin $J$, characterising
the BTZ spacetimes. 

The isometry group of adS3 is usually obtained \cite{HawkingEllis} by
first characterising adS3 as a hyperboloid embedded in $\R^{(2,2)}$
endowed with a flat metric. Its symmetry group is then the subgroup of
the isometry group of the embedding space that leaves the hyperboloid
invariant, namely $SO(2,2)$. This approach gives us only the BTZ
solutions for $M>|J|-1$. This includes all black hole solutions, but
only a region of point particles and overspinning particles. 

More explicitly, one can write the adS3 metric locally in terms of
``cut and paste'' coordinates $(\hat t,\chi,\hat \phi)$ in which the
metric coefficients depend only on $\chi$. The BTZ metrics are then
obtained by identifying $\hat\phi$ with a period smaller or larger
than $2\pi$, and identifying $\hat t$ with a jump across the cut. The
identifications not parameterised by an element of the isometry group
are those in which the period is larger than $2\pi$ (rather than
smaller), or the time jump larger than $2\pi$.

Spinning massless particles are obtained from point particles by
applying an infinite boost and sending the mass to infinity
simultaneously. They correspond to a non-trivial identification of
adS3 across a null plane but have $M=-1$ and $J=0$ like
adS3 itself. 

Restricting ourselves to point particles with $|J|-1\le M<-|J|$, or
the vacuum exterior of compact objects in that parameter range, or the
massless equivalents of point particles, we take an algebraic approach
to the two-body problem in 2+1 dimensions with $\Lambda<0$ in which
each object is obtained by an identification under isometry, and
the resulting combined object is obtained by multiplying the two
elements of the isometry group. We now summarise previous work using
this approach.

Time-symmetric initial data for two massive point particles, two black
holes, or one black hole and one massive point particle were
constructed in \cite{Steif96}. Complete solutions describing two
massless non-spinning particles without and with impact parameter were
constructed in \cite{M99,HM99}, and two or more massive non-spinning
particles colliding at one point in \cite{L16a,L16b}. With non-zero
impact parameter, above a certain energy threshold the spacetime
contains closed timelike curves (from now on, CTCs), in analogy with
the Gott time machine \cite{Gott91}. However, if the boundary of the
region of CTCs is considered as the physical singularity, this
spacetime instead represents the creation of a rotating black hole
from two particles \cite{HM99}.

In this paper, we generalise the algebraic approach of
\cite{Steif96,M99,HM99,L16a,L16b} to computing the total mass and spin
of the most general two-point-particle setup, allowing for arbitrary rest
masses (including zero) and spins, and arbitrary center-of mass
relative momentum and impact parameter. We do not, however, generalise
the explicit spacetime constructions of \cite{M99,HM99}. Rather, we
rely on the beautiful constructions in \cite{HM99} as providing a
sufficiently general example to argue that a black hole forms from two
point particles if and only if the total mass $M$ and spin $J$ obey
the inequalities $M\ge 0$ and $|J|\le M$ that characterise black holes.

We begin with parameter counting. Each compact object surrounded by
vacuum has a rest mass (which can be zero), a spin (which can be
zero), an initial position and an initial momentum. Hence the two-body
initial data have 12 parameters. Six of these correspond to rotating,
boosting and translating the two-body combined object, and so in the
absence of any further objects in the universe they are pure
gauge. The other six parameters of the initial data are
gauge-invariant. One can take them to be the two rest masses, the two
spins, the impact parameter measured in the rest frame, plus one
more. For two massless nonspinning particles, we take this last
parameter to be their energies in the rest frame (where they are
equal). For two massive particles, we take it to be their total
relative rapidity. For one massless and one massive particle we use
the rapidity of the massive particle and the energy of the massless
particle, both with respect to the rest frame.

In Sec.~\ref{section:BTZ}, based on
\cite{DeserJackiwTHooft84,BTZ92,BHTZ93,C95,MiskovicZanelli2009}, we
review in explicit coordinates how metrics describing black holes,
point particles and overspinning particles can be characterised as
identifications of adS3. We discuss their spacetime structure and the
role of closed timelike curves (CTCs). 

In Sec.~\ref{section:SL2R}, based on
\cite{M99,HM99,BHTZ93,C95,MiskovicZanelli2009}, we review a group
theoretical approach complementary to the metric approach of
Sec.~\ref{section:BTZ}, where adS3 is identified with the group
manifold $SL(2,\R)$ and the isometry group of adS3 is represented by
$SL(2,\R)\times SL(2,\R)/{\mathbb Z}^2$ acting by conjugation, rather
than by $SO(2,2)$ acting on $\R^{(2,2)}$. The two groups are related
in Appendix~\ref{appendix:SO22SL2R}. We also point out that either
group theory approach gives only BTZ solutions with $|J|<M+1$.
Appendix~\ref{appendix:SO22overspinning} relates the coordinate and
isometry group treatments of overspinning
particles. 

Appendix~\ref{appendix:Lambda0} considers the limit $\Lambda=0$ from
both the coordinate and isometry group approach. We point out that
spinning point particles can be defined in two ways, and have in fact
been defined differently for $\Lambda=0$ and $\Lambda<0$ in the
literature. We give a BTZ-like metric for point particles. We also
give an exact expression for the total angular momentum of a pair of
particles in the $\Lambda=0$ case that is missing from
\cite{DeserJackiwTHooft84}.

In Sec.~\ref{section:singleparticle} we review, based on
\cite{M99,HM99,L16a,L16b}, how the isometries corresponding to massive
non-spinning point particles (the ``particle generators'') can be
obtained from their geodesics. Massless particles are obtained in the
limit of infinite boost and vanishing rest mass. In an alternative and
more general approach, we obtain the generators for spinning massive
point particles on arbitrary trajectories (for a single object in the
universe, the trajectory is still pure gauge) by boosting and
translating the generators of a single massive point particle sitting
at the centre of the coordinate system. Again massless particles are
obtained as a limit. The nonspinning massive particle case serves as
check on this calculation. 

In Sec.~\ref{section:twoparticles} we create systems of two objects by
multiplying their generators. We explain in detail how the two
different product orders correspond to different gauges, and how to
define gauge-invariant physical quantities, such as the relative
rapidity and the impact parameter of the particles in the rest frame
of the collision. Appendix~\ref{appendix:effectiveparticle} gives a simple
example of how the product generators define an effective particle and
how the order of multiplication corresponds to a gauge choice. 

In Sec.~\ref{section:finalstate} we obtain the total mass
$M_\text{tot}$ and spin $J_\text{tot}$ of the two-particle system,
based on the product generators. Motivated by the work of Holst and
Matschull \cite{HM99}, we {\em conjecture} that a black hole (with
mass $M_\text{tot}$ and spin $J_\text{tot}$) forms if and only
$|J_\text{tot}|<M_\text{tot}$. Otherwise, they define an effective
point particle or overspinning particle characterising the metric
outside of both objects. If the two particles collide they could also form a
single real particle. Appendix~\ref{appendix:Steif} summarises the
work of Steif \cite{Steif96} on the time-symmetric case, which serves as
a check on our results. 

The algebraic shortcut approach for calculating $M_\text{tot}$ and
spin $J_\text{tot}$ of a two-body system has previously been used in
\cite{BS00}. There, they are expressed as functions of three
parameters, but it remains unclear which three-dimensional subspace of
the six-dimensional parameter space of the general two-body system is
being examined.

Sec.~\ref{section:conclusions} contains our conclusions and a list of
open questions. 


\section{Metric description of adS3 and its identifications}
\label{section:BTZ}


\subsection{AdS3}


We consider the three-dimensional Einstein equations
\begin{equation}
\label{EE}
G_{ab}+\Lambda\, g_{ab}=8\pi T_{ab}.
\end{equation}
We use units where $c=G=1$. Note that in 2+1 Newton's constant $G$
has units of 1/mass, so in such units mass is dimensionless and
angular momentum has dimension of length.

We write the negative cosmological constant as
\begin{equation}
\Lambda=:-{1\over\ell^2}<0.
\end{equation}
From now on and unless otherwise stated, we measure length and time in
units of $\ell$, so that we can set $\ell=1$, and all quantities
become dimensionless. $\ell$ can be reinstated from dimensional
analysis, assuming that, still with $c=G=1$, $\ell$ itself and our
coordinates $t$, $r$, $\rho$, $\chi$, $x_i$ and spin $J$ have
dimension length, while the angle $\phi$ and mass $M$ are
dimensionless.

Any solution of the Einstein equations (\ref{EE}) with $T_{ab}=0$ is
locally isometric to adS3. This can be characterised as the
hyperboloidal hypersurface
\begin{equation}
\label{hyperboloid}
  -x_3^2-x_0^2+x_1^2+x_2^2=-1,
\end{equation}embedded in $\R^{(2,2)}$ and endowed with the metric induced by the
flat metric
\begin{equation}
\label{ads3uvxymetric}
ds^2 = -dx_3^2-dx_0^2+dx_1^2+dx_2^2
\end{equation}
on $\R^{(2,2)}$ \cite{HawkingEllis}. The entire hypersurface
(\ref{hyperboloid}) can be parameterised as
\begin{subequations}
\label{adS3parameterisation}
\begin{eqnarray}
x_3&=& \cosh{{\chi}}\cos{t}, \\
x_0&=& \cosh{{\chi}}\sin{t}, \\
\label{x1def}
x_1&=& \sinh{{\chi}}\cos{\phi}, \\
\label{x2def}
x_2&=& \sinh{{\chi}}\sin{\phi},
\end{eqnarray}
\end{subequations}
where the coordinate ranges are
\begin{equation}
\label{chitphiadS3ranges}
0\le{\chi}<\infty, \quad 0\le{t}<2\pi, \quad 0\le{\phi}<2\pi.
\end{equation}
In these coordinates the induced metric becomes 
\begin{equation}
\label{adS3metric}
ds^2 = -\cosh^2{{\chi}}\, d{t}^2+d\chi^2 +\sinh^2{{\chi}}\, d{\phi}^2.
\end{equation} 
The maximal analytic extension of adS3 is obtained by dropping the
periodicity of ${t}$, thus taking the universal cover of the
original, periodic, version. Each slice of constant $t$ is the
hyperpolic 2-plane (from now on ${\mathbb H}^2$), which has constant negative
curvature. For clarity, we will from now on refer to the periodic
version as padS3 and the extended version as eadS3. 

With the new radial coordinate
\begin{equation}
\label{rdef}
r:=\sinh{{\chi}}=\sqrt{x_1^2+x_2^2}, 
\end{equation}
the metric of eadS3 can be written in the alternative form
\begin{equation}
\label{adS3metric2}
ds^2 = -\left(1+r^2\right)\, d{t}^2
+\left(1+r^2\right)^{-1}\,dr^2 + r^2\, d{\phi}^2,
\end{equation} 
with coordinate ranges 
\begin{equation}
\label{trphiranges}
0\le r<\infty, \quad -\infty<{t}<\infty, \quad 0\le{\phi}<2\pi.
\end{equation}

A third useful form of the eadS3 metric is obtained by defining the
radial coordinate
\begin{equation}
\rho:= \tanh{\chi\over 2} \qquad 0\le \rho<1, 
\end{equation}
which implies
\begin{equation}
r={2\rho\over 1-\rho^2}.
\end{equation}
The inverse can be written as
\begin{equation}
\label{rhofromr}
\rho={r\over \sqrt{r^2+1}+1}.
\end{equation}
The metric becomes
\begin{equation}
\label{adS3metric3}
ds^2={4\over (1-\rho^2)^2}\left(-{(1+\rho^2)^2\over 4}d{t}^2
+d\rho^2+\rho^2\,d{\phi}^2\right).
\end{equation}
In these coordinates, each ${\mathbb H}^2$ slice of constant ${t}$ is
represented in conformally flat form, that is as the Poincar\'e disk.
Timelike null infinity $r=\infty$ is now represented by the boundary
$\rho=1$. The conformal diagram of eadS3 is a cylinder. In the
resulting spacetime picture in coordinates $(t,\rho,\phi)$, lightcones are
isotropic in $\phi$ but twice as wide at
the centre $\rho=0$ as at the boundary $\rho=1$.

In a fourth coordinate system on eadS3, we introduce the tortoise
radius
\begin{equation}
r_*:=\int{dr\over 1+r^2}=\tan^{-1} r,
\end{equation}
with range
\begin{equation}
0\le r_*<{\pi\over 2},
\end{equation}
or equivalently
\begin{equation}
\sinh\chi=\tan r_*\quad \Leftrightarrow \quad
\cosh\chi={1\over\cos r_*} ,
\end{equation}
to write the metric (\ref{adS3metric}) as 
\begin{equation}
\label{eadS3rstar}
ds^2={1\over \cos^2r_*}\left(-d
t^2+dr_*^2+\sin^2r_*\,d\phi^2\right).
\end{equation}
The conformal spatial metric is now that of one half of a 2-sphere,
representing the hyperbolic plane as the Klein disk. The lightcones
are at 45 degrees in the radial direction, but are no longer
isotropic.

Rotating the coordinates on the 2-sphere via
\begin{eqnarray}
\sin r_*\cos\phi &=&\cos\theta, \\
\sin r_*\sin\phi &=&\sin\theta\cos\varphi, \\
\cos r_* &=&\sin\theta\sin\varphi, 
\end{eqnarray}
(any two of these three equations are independent) gives
\begin{equation}
\label{eadS3Klein}
ds^2={1\over \sin^2\theta\sin^2\varphi}\left(-d
t^2+d\theta^2+\sin^2\theta\,d\varphi^2\right).
\end{equation}
Whereas the obvious family of null geodesics of (\ref{eadS3rstar}),
$r_*=t$ at constant $\phi$, form a null cone, all
intersecting at the point $r_*=t=0$, the equally obvious family of
null geodesics of (\ref{eadS3Klein}), $\theta=t$ at constant
$\varphi$, are parallel and meet only at the points $\theta=t=0$
and $\theta=t=\pi$ on the conformal boundary: they form the adS3
equivalent of a null plane \cite{HM99}.


\subsection{BTZ black holes}


The BTZ metric can be obtained by making an axistationary ansatz for
the Einstein equations (\ref{EE}) in vacuum. It is \cite{BTZ92}
\begin{equation}
\label{BTZmetric}
ds^2=-f\,d t^2+f^{-1}\,d r^2+r^2(d\phi+\beta dt)^2,
\end{equation}
where 
\begin{equation}
\label{fbetadef}
f:=-M+{r^2}+{J^2\over 4r^2}, \qquad
\beta:=-{J\over 2r^2}.
\end{equation}
$M$ is dimensionless but $J$ has dimension length if we do not use
units where $\ell=1$.  Clearly the case $M=-1$ and $J=0$ is the eadS3
metric in the form (\ref{adS3metric2}). For any $M$ and $J$, the
surfaces of constant $t$ are spacelike at sufficiently large $r$. The
coordinates have the ranges (\ref{trphiranges}).  As a matter of
convention, throughout this paper spacetime points with coordinates
$(t,r,\phi)$ and $( t,r,\phi+2\pi)$ are always identified. We write
such identifications as
\begin{equation}
\label{tidentify}
(t,r,\phi)\sim(t,r,\phi+2\pi).
\end{equation}
We note already that the same identification will look different in
the coordinates ${\hat\phi}$ and ${\hat t}$ introduced later.

We focus first on the parameter range $M>0$ with $0\le|J|\le M$, for
which the BTZ metric represents a black hole
\cite{BTZ92,BHTZ93,C95}. We define the dimensionless parameters
\begin{equation}
\label{lambdaplusdefs}
\lambda_{+\pm}:=\sqrt{M\pm J}
\end{equation}
and their linear combinations
\begin{equation}
\label{spmdef}
s_\pm:={1\over 2}(\lambda_{++}\pm\lambda_{+-}).
\end{equation}
We note for later the identities
\begin{eqnarray}
\label{BTZmass}
M&=&s_+^2+s_-^2, \\ 
\label{BTZangmom}
J&=&2s_+s_-, \\
\sqrt{M^2-{J^2}}&=&s_+^2-s_-^2.
\end{eqnarray}
$s_-$ has the same sign as $J$, while $s_+>0$. The metric coefficient
$f$ has zeros at $r^2=s_\pm^2$. We define $r_+:= s_+$ and
$r_-:=|s_-|$. They obey $0\le r_-\le r_+$. The Killing vector
$\partial/\partial t$ is timelike in the outer region $r>r_+$,
spacelike in the middle region $r_-<r<r_+$ and again timelike in the
inner region $r<r_-$.

As in the Kerr solution in 3+1 spacetime dimensions, Kruskal
coordinates can be constructed to show that $r=r_+$ is an event
horizon separating the outer and middle regions, and $r=r_-$ a Cauchy
horizon separating the middle and inner regions \cite{BHTZ93}. 

In the non-spinning case $J=0$, we have $r_+=\sqrt{M}$ and $r_-=0$, so
the inner region does not exist. In the extremal case $M=|J|>0$, we
have $r_+=r_-=\sqrt{M/2}$ and the middle region does not exist. We
have referred here to ``the'' inner, middle and outer region, event
horizon and Cauchy horizon, but in the maximally extended non-rotating
black hole solutions there is a ``left'' as well as a ``right'' outer
region, and in the spinning ones all these are repeated to the past
and future, see \cite{BTZ92} or \cite{C95} for conformal diagrams.

In the subextremal case $|J|<M$, the BTZ black hole solution
(\ref{BTZmetric}) can be locally identified with the hyperboloid
(\ref{hyperboloid}), where formulas for the $x_\mu$ are given in the
first three columns of Table~\ref{table:BTZ}  in terms of
intermediate coordinates $({\hat t},{\chi},{\hat\phi})$. Those in turn
are given in terms of the BTZ coordinates $(t,r,\phi)$ by 
\begin{subequations}
\label{hatdefsBH}
\begin{eqnarray}
\label{thatdefBH}
\chi&=& \begin{cases}
\cosh^{-1}\sqrt{\alpha}, & \text{outer region} \\
\sin^{-1}\sqrt{\alpha}, & \text{middle region}\\
-\sinh^{-1}\sqrt{-\alpha}, & \text{inner region}
\end{cases}, \\
\label{thatdefBH}
{\hat t} &=& s_+t -s_-\phi, \\
\label{phihatdefBH}
{\hat\phi} &=& s_+\phi -s_-t,
\end{eqnarray}
\end{subequations}
where we have introduced shorthand \cite{C95}
\begin{equation}
\label{alphadef}
\alpha := {r^2-r_-^2\over r_+^2-r_-^2}
={{r^2}-s_-^2\over s_+^2-s_-^2}.
\end{equation}
It is straightforward to verify that in each of the three regions the
metric in $(t,r,\phi$) induced by (\ref{ads3uvxymetric}) is indeed the
BTZ metric (\ref{BTZmetric}).  Note that the embeddings for the three
regions of a black hole given in Table~\ref{table:BTZ} differ from
those given in \cite{BHTZ93,C95} by a gauge transformation. The gauge
here has been chosen so that for nonspinning black holes the
intersection of the plane $x_3=0$ with the hyperboloid is a moment of
time symmetry, parameterised as $\hat t=0$, and that their generators
(see below) then obey $u=v$. An explicit transformation
from the BTZ metric to the adS3 metric (in the Poincar\'e coordinates)
in the extremal case $|J|=M>0$ is given in \cite{BHTZ93}.

\setlength{\tabcolsep}{8pt}
\renewcommand{\arraystretch}{1.2}
\begin{table*}
\begin{tabular}{r|c|c|c|c|c}
& outer region & middle region & inner region & point particle &
  overspinning particle \\
\hline 
& $r_+< r<\infty$ & $r_-< r< r_+$ & $0< r< r_-$ &
$0<r<\infty$ & $0<r<\infty$ \\
& $0< {\chi} <\infty$ & 
$0< {\chi} < {\pi\over 2}$ &
$\chi_{\rm bh}<{\chi}< 0$ & 
$\chi_{\rm pp}<{\chi}<\infty$ & $\chi_{\rm os}<{\chi}<\infty$ \\
$x_3$ & $- \sinh{{\chi}} \, \sinh {\hat t}$ & 
$-\cos{{\chi}}\,\cosh{\hat t}$ & 
$-\cosh{{\chi}}\,\cosh {\hat t}$ &
$\cosh{{\chi}}\cos{\hat t}$ & 
$-\cosh{{\chi}}\cosh{\tilde\phi} \sin{\tilde t}
+\sinh{{\chi}} \sinh{\tilde\phi}\cos{\tilde t}$ 
\\
$x_0$ & $-\cosh{{\chi}} \,\cosh {\hat\phi}$ 
& $- \sin{{\chi}} \,\cosh{\hat\phi}$ &
$ \sinh{{\chi}} \, \sinh {\hat\phi}$ &
$\cosh{{\chi}} \sin{\hat t}$ & 
$\cosh{{\chi}}\cosh{\tilde\phi}\cos{\tilde
  t}+ \sinh{{\chi}} \sinh{\tilde\phi} \sin{\tilde t}$
\\
$x_1$ & $\cosh{{\chi}} \, \sinh {\hat\phi}$ & 
$ \sin{{\chi}} \, \sinh {\hat\phi}$ &
$- \sinh{{\chi}} \,\cosh{\hat\phi}$ &
$ \sinh{{\chi}}\cos{\hat\phi}$ & 
$-\cosh{{\chi}} \sinh{\tilde\phi}\cos{\tilde t}
- \sinh{{\chi}}\cosh{\tilde\phi} \sin{\tilde t}$ \\ 
$x_2$ & $ \sinh{{\chi}} \,\cosh {\hat t}$ & 
$\cos{{\chi}}\, \sinh {\hat t}$ & 
$\cosh{{\chi}}\, \sinh {\hat t}$ &
$ \sinh{{\chi}} \sin{\hat\phi}$ & 
$-\cosh{{\chi}} \sinh{\tilde\phi} \sin{\tilde t}
+\sinh{{\chi}}\cosh{\tilde\phi}\cos{\tilde t}$ \\
\end{tabular}
\caption{\label{table:BTZ} Embedding of the BTZ solution into
  eadS3, in terms of $x^\mu$. The constants $\chi_\text{bh}$,
  $\chi_\text{pp}$ and $\chi_\text{os}$ are defined in
  (\ref{chibhdef}), (\ref{chippdef}) and (\ref{chiosdef}).}
\end{table*}

The induced metric in $({\hat t},{\chi},{\hat\phi})$ is
\begin{subequations}
\label{BTZtchiphimetric}
\begin{eqnarray}
\label{BTZtchiphimetricouter}
ds^2&=&-\sinh^2{{\chi}}\,d{\hat t}^2+d\chi^2+\cosh^2{{\chi}}\,
d{\hat\phi}^2, \\
\label{BTZtchiphimetricmiddle}
ds^2&=&\cos^2{{\chi}}\, d{\hat t}^2-d\chi^2+\sin^2{{\chi}}\,d{\hat\phi}^2, \\
ds^2&=&\cosh^2{{\chi}}\, d{\hat t}^2+d\chi^2-\sinh^2{{\chi}}\,d{\hat\phi}^2
\label{BTZtchiphimetricinner}
\end{eqnarray}
\end{subequations}
for the outer, middle and inner regions, respectively.  These metrics
are therefore alternative local forms of the adS3 metric. They are the
only forms of the adS3 metric that are diagonal and depend on only one
coordinate $\chi$, made unique by the choice $|g_{\chi\chi}|=1$.

We see that $\partial/\partial \hat t$ is the Killing generator of the
event horizon, while $\partial/\partial \hat\phi$ is the Killing
generator of the Cauchy horizon. In BTZ coordinates, the event horizon
and Cauchy horizon generators are $\partial/\partial
t+\Omega_\pm\partial/\partial\phi$ where $\Omega_\pm=J/(2r_\pm^2)$,
respectively.

The range of ${\chi}$ in each of the three regions is given in the
second row of Table~\ref{table:BTZ}, with
\begin{equation}
\label{chibhdef}
\chi_{\rm bh}:=-\sinh^{-1}{|s_-|\over\left(M^2-{J^2}\right)^{1/4}}.
\end{equation}
$\chi=\chi_{\rm bh}$ corresponds to $r=0$ in the inner patch. 

From (\ref{thatdefBH},\ref{phihatdefBH}), the identification
(\ref{tidentify}) in BTZ coordinates is equivalent to
\begin{equation}
\label{hatidentifyBH}
({\hat t},{\chi},{\hat\phi})\sim({\hat t}-2\pi s_-,{\chi},\hat
\phi+2\pi s_+). 
\end{equation}
In particular, the angle $\hat\phi$ has period $2\pi s_+$ (with no
particular significance of $s_+=1$), and the time coordinate ${\hat
  t}$ is identified with a jump
\begin{equation}
\Delta\hat t_{\rm bh}:=-2\pi s_-,
\end{equation}
which vanishes when $J=0$ and has the opposite sign from $J$.  We
shall refer to the intermediate coordinates $(\hat t,\chi\,\hat\phi)$
also as the cut-and-paste coordinates, in contrast to the BTZ
coordinates $(t,r,\phi)$.

With the identification (\ref{hatidentifyBH}), the surfaces of
constant $\hat t$ in the outer metric are wormholes, with a throat of
circumference $2\pi r_+$ located at $\chi=0$. They can be interpreted
as the Killing slicing of the two exterior regions of the Kruskal
metric, with time going forwards in the right outer region, backwards
in the left outer region, and each time slice going through the
2-surface $\chi=0$, equivalent to $x_0=x_1=0$, where the Killing
horizon bifurcates. 


\subsection{Point particles}


We next focus on the parameter range $M<0$, $|J|<-M$. Then the BTZ
metric (\ref{BTZmetric}) has no horizons (zeros of $f$ for real
$r$). Instead the BTZ solution for $M<0$ but $M\ne -1$ represents what
one could either call a naked singularity \cite{BTZ92} or a point
particle \cite{DeserJackiwTHooft92}. We define the two dimensionless
parameters
\begin{equation}
\label{lambdminusdefs}
\lambda_{-\pm}:=\sqrt{-M\pm J},
\end{equation}
and their linear combinations
\begin{equation}
\label{apmdef}
a_\pm:={1\over 2}(\lambda_{-+}\pm\lambda_{--})
\end{equation}
as the equivalent of the BTZ black hole parameters $s_\pm$. They obey
\begin{eqnarray}
\label{particlemass}
-M&=&a_+^2+a_-^2, \\ 
\label{particleangmom}
J&=&2a_+a_-, \\
\sqrt{M^2-{J^2}}&=&a_+^2-a_-^2.
\end{eqnarray}
$a_-$ has the same sign as $J$, while $a_+>0$. 

The identification of the point particle BTZ solution with padS3
written as the hyperboloid (\ref{hyperboloid}) in $\R^{(2,2)}$ was
found in \cite{MiskovicZanelli2009}. (However,
\cite{MiskovicZanelli2009} restrict to $|J|\le -M$ for $-1\le M<0$
only. We do not see why this would be necessary.) This
identification is given here in the second-last column of
Table~\ref{table:BTZ}, with the intermediate coordinates now given in
terms of the BTZ coordinates by
\begin{subequations}
\label{hatdefspp}
\begin{eqnarray}
\label{chitildedef}
{\chi} &=&\cosh^{-1}\sqrt{\alpha} \\
{\hat t} &=& a_+t +a_-\phi, \\
{\hat\phi} &=& a_+\phi+a_-{t}.
\end{eqnarray}
\end{subequations}
Here $\alpha$ is defined by 
\begin{equation}
\alpha={{r^2}+a_+^2\over a_+^2-a_-^2}.
\end{equation}
[This is the same expression as (\ref{alphadef}) if we take into
  account that $s_\pm^2=-a_\mp^2$.]  The range of ${\chi}$ is given in
Table~\ref{table:BTZ}, with 
\begin{equation}
\label{chippdef}
\chi_{\rm pp}:=\cosh^{-1}{a_+\over \left(M^2-{J^2}\right)^{1/4}}.
\end{equation}
$\chi=\chi_{\rm pp}$ corresponds to $r=0$. The expressions for the
$x_{\mu}$ given in Table~\ref{table:BTZ} in the point particle case
are simply (\ref{adS3parameterisation}) with hats on, and so the
induced metric in $({\hat t},{\chi},{\hat\phi})$ is 
\begin{equation}
\label{ppmetric}
ds^2=-\cosh^2\chi\,d\hat t^2+d\chi^2+\sinh^2\chi\,d\hat\phi^2,
\end{equation}
The induced metric in $(t,r,\phi)$ is once again the BTZ metric
(\ref{BTZmetric}).

The identification(\ref{tidentify}) is now equivalent to
\begin{equation}
\label{hatidentifyPP}
({\hat t},{\chi},{\hat\phi})\sim
({\hat t}+2\pi a_-,{\chi},{\hat\phi}+2\pi a_+).
\end{equation}
We can think of this as a wedge cut out with defect angle $2\nu>0$, or
a wedge inserted with excess angle $2\nu<0$, where
\begin{equation}
\label{nuppdef}
\nu:=\pi\left(1-a_+\right),
\end{equation}
and a time jump
\begin{equation}
\label{Deltathatdef}
\Delta{\hat t}_{\rm pp}:=2\pi a_-,
\end{equation}
which has the same sign as $J$, applied when identifying across the
sides of the wedge. Hence we can think of the point particle geometry
as eadS3 with a  timelike conical singularity and time jump. The exception is
$M=-1$ and $J=0$, which gives $a_+=1$ and $a_-=0$.


\subsection{Overspinning particles}
\label{section:overspinningparameterisation}


The $(J,M)$-plane is completed by two disjoint overspinning regions
$|J|>|M|$. There are no horizons, and, as we will see below, if we
restrict to $r\ge 0$ there are no closed timelike curves either. Hence
we can consider the restriction most usefully as a kind of
particle. 

For clarity, in this paper we will call BTZ solutions with $M<0$,
$|J|<-M$ ``point particles'', and BTZ solutions with $|J|>|M|$
``overspinning particles''.  Only the nonspinning particle solutions
are unambiguously ``point particles'', with a singular worldline at
$r=\chi=0$, whereas in both spinning point particle and overspinning
particle solutions a singular worldline at $\chi=0$ is surrounded by a
world tube containing closed timelike curves whose outer boundary is
at $r=0$, see the following Sec.~\ref{section:CTCregion} for
details. Therefore either all (spinning) point particle and
overspinning particle solutions should be considered as ``particles''
or none. We have opted here for the former as the one more consistent
with the established terminology in the literature.

To simplify notation, we now restrict to the case $J>|M|$. The case
$J<-|M|$ can be obtained by flipping the signs of $\hat\phi$ and $J$,
thus also replacing $\lambda_{\pm +}$ with $\lambda_{\pm -}$.

An identification of this class of solutions with padS3 in the form of
(\ref{hyperboloid}) is given in the last column of
Table~\ref{table:BTZ}, with 
\begin{subequations}
\label{hatdefsos}
\begin{eqnarray}
\chi&=&{1\over
  2}\sinh^{-1}\left({2{r^2}-M\over\sqrt{{J^2}-M^2}}\right), \\
{\tilde t}&=& {\lambda_{-+}\over 2}(t+\phi), \\
{\tilde\phi}&=&{\lambda_{++}\over 2}\left(\phi-t\right).
\end{eqnarray}
\end{subequations}
This is derived in Appendix~\ref{appendix:SO22overspinning} (using the
methods of Sec.~\ref{section:SL2R} below and Appendix~\ref{appendix:SO22SL2R}.)
 The value of $\chi$ corresponding to $r=0$ is
\begin{equation}
\label{chiosdef}
\chi_{\rm os}:={1\over
  2}\sinh^{-1}\left({-M\over\sqrt{{J^2}-M^2}}\right).
\end{equation}
The metric in $(t,r,\phi)$ is again the BTZ metric (\ref{BTZmetric}).
The induced metric in intermediate coordinates $({\tilde
  t},\chi,{\tilde\phi})$ is
\begin{subequations}
\label{overspinningmetricthatchiphihat}
\begin{eqnarray}
ds^2&=&-d{\tilde
  t}^2+d\chi^2+d{\tilde\phi}^2+2\sinh {2{\chi}}\,d{\tilde\phi}\,d{\tilde
  t} \\
&=&-\cosh^2{2{\chi}}\,d\tilde t^2+d\chi^2
+\left(d\tilde\phi+\sinh {2{\chi}}\,d\tilde t\right)^2. \nonumber \\
\end{eqnarray}
\end{subequations}
As a byproduct we have found yet another form of writing the metric of
eadS3, see also \cite{BengtssonSandin2006}. Any parameterisation of
the overspinning BTZ metric that represents it as an identification of
adS3 with a shift in $\tilde t$ and a shift in $\tilde\phi$, where
$\partial/\partial \tilde t$ and $\partial/\partial \tilde\phi$ are
commuting Killing vectors of adS3, both before and after the
identification, must be related to this one by a linear recombination
of the coordinates $\tilde t$ and $\tilde\phi$ and a reparameterisation of
the coordinate $\chi$. It is clear that no such reparameterisation can
make the metric diagonal at the same time. 

The identification (\ref{tidentify}) is equivalent to
\begin{equation}
\label{overspinningdefectangleandtimeshiftplus}
({\tilde t},\chi,{\tilde\phi})\sim({\tilde
  t}+\pi\lambda_{-+},\chi,{\tilde\phi}+\pi\lambda_{++})
\end{equation}
To make this look more like the point particle and black hole cases,
we define the alternative cut-and-paste coordinates
\begin{eqnarray}
\hat t&:=&\tilde t-\tilde\phi = b_+t+b_-\phi, \\
\hat\phi&:=&\tilde t+\tilde\phi = b_+\phi+b_-t,
\end{eqnarray}
where we have defined the shorthand parameters
\begin{equation}
b_\pm:={1\over 2}(\lambda_{-+}\pm\lambda_{++}).
\end{equation}
The metric becomes
\begin{equation}
ds^2={1\over 2}\sinh 2\chi(-d\hat t^2+d\hat\phi^2)+d\chi^2
-d\hat t\,d\hat\phi
\end{equation}
(yet another local coordinate system on adS3), and the identification
is
\begin{equation}
({\hat t},\chi,{\hat \phi})\sim({\hat
  t}+2\pi b_-,\chi,{\hat \phi}+2\pi b_+).
\end{equation}
For the prototype overspinning particle $J=1$, $M=0$, this reduces
$({\hat t},\chi,{\hat\phi})\sim({\hat t},\chi,{\hat\phi}+2\pi)$ just
as for adS3 spacetime $M=-1$, $J=0$ and the prototype black hole $M=1$
and $J=0$.


\subsection{Closed timelike curves}
\label{section:CTCregion}


By replacing the coordinate $r$ with $R:=r^2$ in (\ref{BTZmetric}) we
obtain
\begin{equation}
\label{BTZmetricFR}
ds^2=\left(M-{R }\right)\,dt^2
+{dR^2\over F}+R \,d\phi^2-J\,d\phi\,dt,
\end{equation}
where
\begin{equation}
F:=4R^2-4MR+J^2=r^2f(r).
\end{equation}
Hence we have an analytic continuation beyond $r=0$ to negative
$R$. The metric (\ref{BTZmetricFR}) is regular, with regular inverse,
except where $F=0$. These roots occur at
\begin{equation}
R_\pm:={M\pm\sqrt{M^2-J^2}\over 2}.
\end{equation}
For point particles, $R=R_+=-a_-^2<0$ corresponds to the particle
location $\chi=0$, which is a conical singularity. Therefore, the
maximal analytic extension of the spacetime corresponds to
$R_+<R<\infty$. The root $R=R_-$ of $F$ is not physical.  For black
holes or overspinning particles, the maximal analytic continuation of
the spacetime corresponds to the range $-\infty<R<\infty$. For black
holes, $R=-\infty$ is a segment of timelike infinity deep inside the
black hole, see the Penrose diagram in \cite{BHTZ93}.

It is clear from the form (\ref{BTZmetricFR}) of the metric that there
is a smooth closed timelike curve (CTC) through every point of the
spacetime with $R<0$, namely the curve given by constant $t$ and $R
$.  Conversely, it was shown in \cite{BHTZ93} that if spacetime is
restricted to $R >0$ there are no closed differentiable causal curves
at all. We give this argument here for completeness. A differentiable
curve $(t,r,\phi)(\tau)$ is causal if
\begin{equation}
\label{causalcurve}
-f\dot t^2+f^{-1}\dot r^2+r^2\left(\dot\phi+{J\over 2r^2}\dot
t\right)^2\le 0,
\end{equation}
where a dot denotes $d/d\tau$. Causal curves that cross an event
horizon or Cauchy horizon cannot cross it again and so cannot be
closed. Hence it is sufficient to consider curves that remain in $f>0$
and curves that remain in $f<0$. In a spacetime region where $f>0$, we
note that a closed differentiable curve must have one point where
$\dot t=0$. But then there is a contradiction with (\ref{causalcurve})
as long as $r^2>0$. Similarly, in a spacetime region where $f<0$, we
note that a closed differentiable curve must have a point where $\dot
r=0$ to obtain the same contradiction.

It was conjectured in \cite{BHTZ93} that {\em any} field theory matter
falling into a BTZ black hole has divergent stress-energy at $r=0$ and
so turns it into a genuine curvature singularity. Therefore it was
proposed to exclude $r<0$ and consider $r=0$ as the true singularity,
by extension even in the vacuum black hole case.  Excluding the CTC
region for point particles or overspinning particles can be justfied
in the same way. However, if we think of the particle as corresponding
to a singular stress-energy tensor with support on a world line, the
point of view taken (for $\Lambda=0$) in \cite{DeserJackiwTHooft84},
that particle is at $\chi=0$ for point particles and at
$\chi=-\infty$ for overspinning particles, not at $r=0$. A spinning
``particle'' at $r=0$ is really a brane, the point of view taken (for
$\Lambda<0$) in \cite{MiskovicZanelli2009}. For the purpose of our
calculations, it will not be necessary to take a view on this as long
as the $r>0$ regions of the two particles never overlap (or at least
not before before they have fallen into a black hole). It is worth
stressing that the existence of CTCs is independent of the value of
$\Lambda$, see also Appendix~\ref{appendix:Lambda0}.


\section{$SL(2,\R)$ description of the isometries of padS3, and of
  the BTZ solutions}
\label{section:SL2R}



\subsection{padS3 as a Lie group manifold}


The hyperboloid (\ref{hyperboloid}) corresponding to the time-periodic
spacetime padS3 can be mapped to the Lie group $SL(2,\R)$ via the
identification
\begin{equation}
\label{xtomat}
\x={x_3}I+{x_0}\gamma_0+{x_1}\gamma_1
+{x_2}\gamma_2,
\end{equation}
where $I$ is the identity matrix and the
$\gamma$-matrices are 
\begin{equation}
\gamma_0= \left(
  \begin{array}{ c c }
     0 & 1 \\
     -1 & 0
  \end{array} \right), \: \space \gamma_1= \left(
  \begin{array}{ c c }
     0 & 1 \\
     1 & 0
  \end{array} \right), \: \space \gamma_2= \left(
  \begin{array}{ c c }
     1 & 0 \\
     0 & -1
  \end{array} \right).
\end{equation} 
Together these form a basis of real $2\times2$
matrices. The condition (\ref{hyperboloid}) is precisely the condition
$\det \x=1$ for the matrix $\x$ to be an element of the group. 
The inverse of (\ref{xtomat}) is
\begin{eqnarray}
\label{mattox}
x_0&=&-{1\over 2}\tr(\gamma_0\x), \quad
x_1={1\over 2}\tr(\gamma_1\x), \nonumber \\
x_2&=&{1\over 2}\tr(\gamma_2\x), \quad
x_3={1\over 2}\tr(\x).
\end{eqnarray}
We can parameterise the general element of $SL(2,R)$ as
\begin{eqnarray}
\label{g0def}
g_\text{gen}(\zeta,\psi,\varphi)&:=&\cosh\zeta\,(\cos\psi\,I+\sin\psi\,\gamma_0)
\nonumber \\
&&+\sinh\zeta\,(\cos\varphi\,\gamma_1+\sin\varphi\,\gamma_2).
\end{eqnarray}
Taking the inverse of $g$ corresponds to changing the
signs of the coefficients of $\gamma_0$, $\gamma_1$ and
$\gamma_2$. Hence
\begin{equation}
g_\text{gen}(\zeta,\psi,\varphi)^{-1}=g_\text{gen}(-\zeta,-\psi,\varphi).
\end{equation}
Note also that for $g,h\in SL(2,\R)$, $\tr(g^{-1})=\tr g$ and $\tr
gh=\tr hg$. 


\subsection{Isometries}
\label{subsection:isometries}


In the representation of padS3 as $SL(2,\R)$, any isometry $\phi$ can be
represented as
\begin{equation}
\label{ghtransformation}
\phi:\ \x\mapsto \tilde \x:=g^{-1}\x h,
\end{equation}
where $g$ and $h$ are two elements of $SL(2,\R)$, called the left and
right generators of the isometry. $(g,h)$ and $(-g,-h)$ (and only
those two pairs) represent the same isometry. We follow the convention
of \cite{HM99}. (By contrast, \cite{C95} uses the convention $g\x h$.)
The composition of isometries is then given by right matrix
multiplication of the generators $(g,h)$, that is $\phi_1\circ\phi_2$
has generators $(g_1g_2,h_1h_2)$.

The isometry admits fixed points $\x=g^{-1}\x h$ if and only if the
generators are in the same conjugacy class, that is, there exists an
$\x$ such that 
\begin{equation}
\label{ghconjugate}
g=\x h\x^{-1}. 
\end{equation}
In particular, if either $h=I$ or $g=I$ (but not both), $\phi$ is the
left or right action and so acts freely (admits no fixed points).
However, if any fixed points exist, the set of fixed points is
precisely a geodesic \cite{HM99}, which can be interpreted as a
generalised axis of rotation.

The isometry group of padS3 can also be represented as $SO(2,2)$
acting on $\x\in\R^{(2,2)}$ by matrix multiplication. See
Appendix~\ref{appendix:SO22SL2R} for details.


\subsection{Point particles, overspinning particles and black holes as isometries}


As we have already seen, particles and black holes in adS3 can be
represented by identifying the padS3 spacetime under a nontrivial
isometry, that is
\begin{equation}
\label{psiuv}
\psi:\ \x\sim u^{-1}\x v.
\end{equation}
We can also look at the identification $\psi$ (considered as a
physical particle or black hole) under the isometry $\phi$ (considered as a
mere change of coordinate system that leaves the form of the metric
invariant), that is $\tilde\psi:=\phi^{-1}\psi\phi$. In terms of the
$SL(2,\R)$ generators we have
\begin{equation}
g^{-1}\x h\sim g^{-1}u^{-1}\x vh=(g^{-1}ug)^{-1}(g^{-1}\x h)(h^{-1}vh)
\end{equation}
or equivalently 
\begin{equation}
\tilde \x \sim \tilde u^{-1}\tilde \x \tilde v
\end{equation}
where the generators of the same isometry $\psi$, expressed in the new
``coordinate system'' $\tilde \x $ are
\begin{equation}
\label{transformedgenerators}
\tilde u:=g^{-1}ug, \qquad \tilde v:=h^{-1}vh.
\end{equation}
This specifies how the generators $u$ and $v$ of the identification $\psi$
representing a physical particle transform under an independent
isometry $\phi$ with generators $g$ and $h$ representing a coordinate change.

The identification of the spacetime with itself under the isometry
$\psi$ is transitive. Hence with two objects (particle or black hole)
we also have the identifications
\begin{equation}
\label{isometrycomposition}
\psi_2\circ\psi_1:\ \x \sim u_2^{-1}(u_1^{-1}\x v_1)v_2=(u_1u_2)^{-1}\x (v_1v_2)
\end{equation}
and 
\begin{equation}
\psi_1\circ\psi_2:\ \x \sim u_1^{-1}(u_2^{-1}\x v_2)v_1=(u_2u_1)^{-1}\x (v_2v_1)
\end{equation}
We can think of these as the representation of an ``effective
particle''.  We note that we can write
\begin{equation}
\label{transformedgeneratorsbis}
u_1u_2=u_2^{-1}(u_2u_1)u_2, \quad v_1v_2=v_2^{-1}(v_2v_1)v_2.
\end{equation}
Comparing (\ref{transformedgeneratorsbis}) with
(\ref{transformedgenerators}), we see that the product particle
generators taken in the two orders are related by the ``coordinate
transformation'' (\ref{ghtransformation}) generated by
$(g,h)=(u_2,v_2)$.  Hence taking the product isometry in the two
orders corresponds to the same effective particle in two different
coordinate patches. See Appendix~\ref{appendix:effectiveparticle} for
the visualisation of an example with $\Lambda=0$ that illustrates the
effective particle in the two different gauges.

Throughout this paper, we use $\x ,\y,\z$ for points in padS3, $u,v$
for left and right particle generators, and $g,h$ for left and right
generators of a ``coordinate change'', even though they are all
elements of $SL(2,\R)$.

What is the gauge-invariant information in a pair $(u,v)$ of
generators?  The eigenvalues of any matrix are invariant under
conjugation, but since in $SL(2,\R)$ the determinant (product of the
eigenvalues) is always $1$ and there are only two such eigenvalues,
two elements of $SL(2,\R)$ are conjugate if and only if they have the
same trace (sum of the eigenvalues). The product trace $\tr(u_1u_2)$
is also gauge-invariant, and independent of the product order. We will
see that the traces of particle generators encode the rest mass and
spin of the particle, and so the traces of the product generators (in
either order) encode the rest mass and spin of the effective
particle. No other gauge-invariant quantities can be constructed from
$(u,v)$. 

For two spacetime points $\x$ and $\y$ in adS3 linked by a geodesic,
the geodesic distance $d(\x,\y)$ between them is given by
\begin{equation}
{1\over 2}\tr\left(\x^{-1}\y\right)=\begin{cases}\cos d(\x,\y) & \hbox{spacelike,} \\
\cosh d(\x,\y) & \hbox{timelike,} \\
1 & \hbox{null separated.}
\end{cases}
\end{equation}
This is easily verified by considering simple cases and noting that
both the left-hand side and the right-hand side of this equation are
gauge-invariant. 


\subsection{BTZ generators}
\label{section:BTZgenerators}


The isometry $\psi$ of the BTZ black
hole solutions is of the form (\ref{psiuv}) with generators
\begin{eqnarray}
\label{BTZBHgenerators}
\left.\begin{aligned}u_{\rm bh}\\v_{\rm bh}\end{aligned}\right\}
&=&-\cosh\beta_\mp\, I
-\sinh\beta_\mp\, \gamma_2 \nonumber \\
&=&-\cosh\left(\pi\sqrt{M\mp J}\right)\, I
-\sinh\left(\dots\right)\, \gamma_2 
\end{eqnarray}
for $M\ge 0$, $|J|\le M$.  We have made an arbitrary choice of
overall sign, such that the generators of padS3 are $I$ [compare the
  limit $M=-1$, $J=0$ of (\ref{BTZparticlegenerators}) below.] To
avoid writing factors of $\pi$, we have introduced the shorthands
\begin{equation}
\beta_\pm:=\pi\lambda_{+\pm}.
\end{equation}
From Table~\ref{table:BTZ} and
(\ref{thatdefBH}-\ref{phihatdefBH}) we can verify explicitly that
(\ref{psiuv}) with (\ref{BTZBHgenerators})
acts on the outer, middle and inner region of the BTZ black hole
spacetime in the same way as $\phi\to
\phi+2\pi$ in the BTZ coordinates $(t,r,\phi)$. However, the
$SL(2,\R)$ picture covers the identifications in all three regions at
once.

The isometry $\psi$ of the BTZ spinning point particle solutions is of the form
(\ref{psiuv}) with generators
\begin{eqnarray}
\label{BTZparticlegenerators}
\left.\begin{aligned}u_{\rm pp}\\v_{\rm pp}\end{aligned}\right\}
&=&\cos\nu_\mp\, I
+\sin\nu_\mp\, \gamma_0 \nonumber \\
&=&-\cos\left(\pi\sqrt{-M\pm J}\right)\, I
+ \sin\left(\dots\right)\, \gamma_0.
\label{pptracesbis}
\end{eqnarray}
To avoid writing factors of $\pi$,
we have introduced the shorthands
\begin{equation}
\label{nupmdef}
\nu_\pm:=\pi(1-\lambda_{-\mp})=\nu\pm{\Delta\hat t_{\rm pp}\over 2}
\end{equation}
We can again verify explicitly that this acts on the BTZ black hole
spacetime in the same way as $\phi\to
\phi+2\pi$ in the BTZ coordinates. 

The BTZ generators for overspinning particles with $J>|M|$ are
\begin{equation}
\label{overspinningplus}
u_{J>|M|}=u_{\rm pp}, \quad v_{J>|M|}=v_{\rm bh},
\end{equation}
and for $J<|M|$ they are
\begin{equation}
\label{overspinningminus}
u_{J>|M|}=u_{\rm bh}, \quad v_{J>|M|}=v_{\rm pp}.
\end{equation}
The coordinate formulas in
Sec.~\ref{section:overspinningparameterisation} were actually found by
the methods of Appendix~\ref{appendix:SO22overspinning} from the
generators (\ref{overspinningplus},\ref{overspinningminus}).

Put differently, in all parts of the $(J,M)$-plane, we choose
$u=u_{\rm bh}$ if $M-J>0$ and $u=u_{\rm pp}$ if $M-J<0$, and we choose
$v=v_{\rm bh}$ if $M+J>0$ and $v=v_{\rm pp}$ if $M+J<0$.

The generators of the $M=J=0$ spacetime appear to be $u=v=-I$, but
this is not the correct limit. From (\ref{hatidentifyBH}), we see that
as $s_+\to 0$, the fundamental domain between the two surfaces
$\hat\phi=0$ and $\hat\phi=2\pi s_+$ disappears. The correct way of
taking the limit is to apply a boost at the same time
\cite{BHTZ93}. The two identification surfaces then do not approach
each other uniformly as the limit is taken, but touch only at one end
(infinity) while staying apart at the other end.

 
\subsection{$M$ and $J$ from the traces of the isometry generators}


The traces of the generators are related to each other in the four
segments of the $(J,M)$-plane by analytic continuation. We have
\begin{subequations}
\label{Tuvanalytic}
\begin{eqnarray}
\label{Tuanalytic}
Tu&=&-\cosh\pi\sqrt{M-J}=-\cos\pi\sqrt{-M+J}, \\
\label{Tvanalytic}
Tv&=&-\cosh\pi\sqrt{M+J}=-\cos\pi\sqrt{-M-J}
\end{eqnarray}
\end{subequations}
for the BTZ generators everywhere in the $(J,M)$ plane. By contrast,
there is no direct analytic continuation of the entire generators. To
obtain one, one would have to apply a gauge transformation to the
generators in order to transform a curve in $SL(2,\R)\times SL(2,\R)$
of generators that turns a corner at $J=\pm M$, with either $u=-I$
or $v=-I$, into a smooth one.

\begin{figure}
\includegraphics[scale=0.65, angle=0]{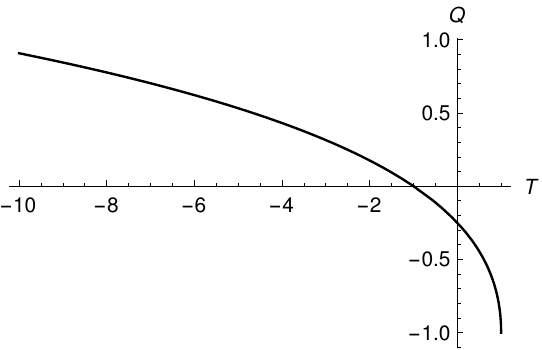} 
\caption{A plot of ${\cal Q}(T)$. It is defined for $T\le 1$. Note
  ${\cal Q}(-1)=0$,
  ${\cal Q}(0)=-1/4$ and ${\cal Q}(1)=-1$.
\label{figure:fig5}}
\end{figure}

\begin{figure}
\includegraphics[scale=0.65, angle=0]{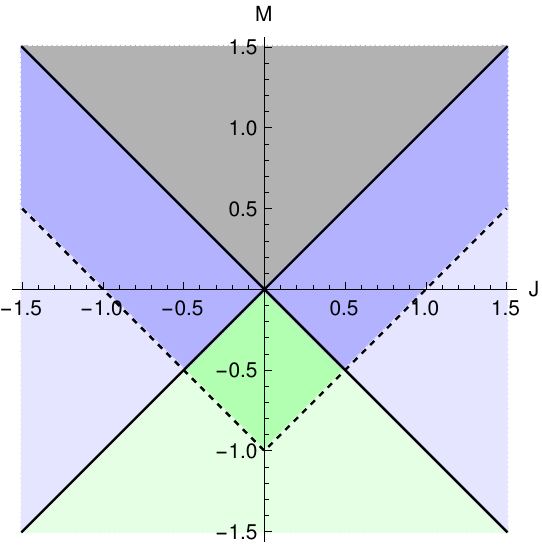} 
\caption{A plot of the $(J,M)$ plane, showing the black hole region
  (black), region of point particles that can be represented using the
  $SL(2,\R)$ approach (dark green) and cannot (light green), and regions of
  overspinning particles that can be represented using the $SL(2,\R)$
  approach (dark blue) and cannot (light blue). The two objects considered
  in this paper take parameters in the dark green region.
\label{figure:JMplane}}
\end{figure}

We define the function
\begin{equation} 
{\cal T}(Q):=-\cosh\pi\sqrt{Q}=-\cos\pi\sqrt{-Q}
\end{equation}
which is complex-analytic in the entire $Q$ plane, and in particular
is real-valued and real-analytic on the real line $-\infty<Q<\infty$. For
$-\infty<T\le 1$, ${\cal T}(Q)$ has the real analytic inverse
\begin{equation}
\label{calQdef}
{\cal Q}(T):={[\cosh^{-1}(-T)]^2\over \pi^2}=-{[\cos^{-1}(-T)]^2\over \pi^2}.
\end{equation}
The function ${\cal Q}(T)$ is shown in Fig.~\ref{figure:fig5}. 
We can then write
\begin{subequations}
\begin{eqnarray}
\label{Tuanalyticbis}
Tu&=&{\cal T}(M-J), \\
\label{Tvanalyticbis}
Tv&=&{\cal T}(M+J)
\end{eqnarray}
\end{subequations}
for all real values of $M$ and $J$, or equivalently
\begin{subequations}
\label{MJfromTuTv}
\begin{eqnarray}
\label{MJanalytic1}
M-J&=&{\cal Q}(Tu), \\
\label{MJanalytic2}
M+J&=&{\cal Q}(Tv).
\end{eqnarray}
\end{subequations}
This bijection between $(u,v)$ and $(J,M)$ is defined only for
$Tu,Tv\le 1$, and $|J|\le M+1$. Fig.~\ref{figure:JMplane} shows the
regions in the $(J,M)$ plane representing black holes, point particles
and overspinning particles, and where these can be represented using
the group theory approach. 

\subsection{Applicability of the isometry group approach}


We have seen that the approach of identifying padS3 or eadS3 under an
isometry of padS3 covers all the BTZ black hole solutions, but point
particles and overspinning particles only for $|J|<M+1$. In other
words, we can only represent point particles in parameter region
\begin{equation}
-1+|J|<M<-|J|
\end{equation}
(the dark green region in Fig.~\ref{figure:JMplane}), and overspinning
particles in the parameter region
\begin{equation}
-|M|<|J|<-1+M
\end{equation}
consisting of two disjoint parts $J>0$ and $J<0$ (the dark blue region
in Fig.~\ref{figure:JMplane}). The BTZ solutions we miss are those
that require identifications
\begin{equation}
\label{myidentification}
(\hat t,\chi,\hat\phi)\sim (\hat t+\Delta\hat
  t,\chi,\hat\phi+\Delta\hat\phi)
\end{equation}
with $|\Delta\hat t|>2\pi$ or $\Delta\hat\phi>2\pi$ or both, whereas
both have period $2\pi$ in the metric (\ref{ppmetric}) of
padS3. Unwinding $\hat t$ gives rise to the eadS3 solution, but
already we are not aware of a parameterisation of the isometry group
of eadS3 that allows easy multiplication. Further unwinding $\hat
\phi$ produces a spacetime that we may call uadS3, with a branch point
at $\chi=0$. Its isometry group is presumably the universal cover of
$SL(2,\R)\times SL(2,\R)$, but this is not a matrix group and so does
not allow for easy multiplication.

At this point, we seem to be facing a technical obstacle that we have
not managed to overcome. However, one may wonder if the the BTZ
solutions we are missing actually arise as exteriors of physical
compact objects. 

This question is suggested by the example of fluid stars. All rigidly
rotating perfect fluid stars, for arbitrary equation of state, have
been constructed explicitly in \cite{2+1rotstars}, building on the
earlier work \cite{Cataldo}. (The construction is formally for
barotropic equations of state, but in an axistationary solution, where
everything depends only on $r$, a stratified equation of state cannot
be distinguished from a barotropic one). For any barotropic equation
of state that has sound speed no greater than the speed of light and
admits solutions with finite mass, stars with exterior metrics of all
three types (point particle, black hole and overspinning particle)
exist. These precisely fill the region $|J|< M+1$ with $M>-1$ of
BTZ parameter space that is characterised by isometries of padS3.


\section{A spinning point particle on an arbitrary trajectory}
\label{section:singleparticle} 


\subsection{Timelike geodesics}


The world line $r=0$ of a point particle given by the BTZ metric
corresponds to the world tube $\chi=\chi_\text{pp}$ surrounding the
geodesic $\chi=0$ in the metric (\ref{ppmetric}). More generally, the
world line of any freely falling compact object corresponds to a world
tube surrounding a geodesic of eads3. Hence we begin by calculating
the timelike geodesics of eadS3.

Since the metric ($\ref{adS3metric})$ is independent of both
$\phi$ and $t$, both $\partial_t$ and $\partial_{\phi}$ are Killing
vectors, giving rise to the conserved energy (per rest mass)
\begin{equation}
E:=-u_{a}(\partial/\partial_t)^a=\cosh^2\chi\ \dot{t}
\end{equation}
and angular momentum (per rest mass)
\begin{equation}
L:=u_{a}(\partial/\partial_{\phi})^a=\sinh^2\chi\ \dot{\phi}. 
\end{equation}
Here a dot denotes $d/d\tau$ and $\tau$ is the proper time. The
normalisation condition $u_au^a=-1$ becomes a nonlinear first-order
differential equation for $\chi(\tau)$, namely
\begin{equation}
\label{timelikegeo} 
\dot\chi^2 + \frac{L^2}{\sinh^2\chi}-\frac{E^2}{\cosh^2\chi}=-1.
\end{equation}
In the coordinate $r$ defined in Eq.~(\ref{rdef}) this becomes
\begin{equation}
\label{rpotential}
\dot r^2=E^2-\left(1+{1\over r^2}\right)L^2-(1+r^2).
\end{equation}
Taking the square root, we obtain a separable ODE that can be
integrated in closed form to obtain
\begin{equation}
\label{chitl}
\chi(\tau)=\sinh^{-1}\left(\sqrt{C - D \cos 2\tau}\right),
\end{equation}
where we have defined the temporary shorthands
\begin{equation}
C:=\frac{E^2-L^2-1}{2}, \qquad D:=\sqrt{C^2-L^2}.
\end{equation}
Without loss of generality, we have fixed the origin of $\tau$ so that
the closest approach to the ``central'' world line $\chi=0$ is at
$\tau=n\pi$ for integer $n$. The proper distance at these moments, 
measured along surfaces of constant $t$ (and in units of $\ell$) is
\begin{equation}
\label{impactparam}
\delta:=\sinh^{-1}\sqrt{C-D}.
\end{equation}  
By integrating the definitions of $E$ and $L$ we then obtain $t(\tau)$
and $\phi(\tau)$.

We define the Lorentz factor $\Gamma$, rapidity $\gamma$ and
2-velocity $v$ of the geodesic with tangent vector $u^a$, with
respect to the observers $n^a$ normal to the slices of constant $t$, by
\begin{equation}
\Gamma:=\cosh\gamma:={1\over \sqrt{1-v^2}}:=-u^a n_a,
\end{equation}
where $u_au^a=n_an^a=-1$.
We note for later use that $v=\tanh \gamma$.
With $n_a=-\cosh\chi\,(dt)_a$, we find
\begin{equation}
\Gamma(\tau)={E\over\cosh\chi(\tau)}.
\end{equation}
We now define the constant parameter
\begin{equation}
z:=\gamma(0)=\cosh^{-1}\left({E\over\cosh\delta}\right).
\end{equation}
In terms of the parameters $\delta$ and $z$, 
we then have
\begin{equation}
E=\cosh z\cosh\delta, \quad L=\sinh z\sinh\delta,
\end{equation}
and the geodesics are given by
\begin{subequations}
\begin{eqnarray}
\label{chitz}
\chi(\tau)&=&\sinh^{-1}
\sqrt{\sin^2\tau\sinh^2z+\cos^2\tau\sinh^2\delta}, \\
\label{ttlz}
t(\tau) &=&\tan^{-1}\left({\cosh z\tan\tau\over\cosh\delta}\right)+t_0, \\
\label{phitz}
\phi(\tau)&=&\tan^{-1}\left({\sinh z\tan\tau\over\sinh\delta}\right)+\phi_0,
\end{eqnarray}
\end{subequations}
where $t_0$ and $\phi_0$ are arbitrary integration constants. We can
now allow both $z$ and $\delta$ to have either sign. If we think about
geodesics in terms of initial data at, say, $t=0$, the free data are
the initial position and 2-velocity, requiring $4$ parameters. These
can be mapped to our parameters $t_0$, $\phi_0$, $\delta$ and $z$, up
to $t_0\to t_0+2n\pi$.

$\chi(\tau)$ is periodic with period $\pi$, and take values between
$|\delta|$ and $|z|$, either of which could be the larger one of the
two. We interpret the inverse $\tan^{-1}$ so that $t(\tau)$ always
increases monotonically, and $\phi(\tau)$ increases (decreases)
monotonically if $\delta$ and $z$ have the same (opposite) sign. When
$\tau$ has increased by $2\pi$, $t$ has increased by $2\pi$, $\phi$
has changed by $2\pi$ or $-2\pi$, and $\chi$ has gone through two
periods.

Looking back, we see that (\ref{rpotential}) is formally identical to
the effective radial equation of motion of a Newtonian particle with
energy per mass $C$, angular momentum per mass $L$, and an
attractive central force per mass equal to the radius $r$ (like the
force of a spring).  Hence it is not surprising that in terms of $x_1$
and $x_2$ the timelike geodesics are the ellipses
\begin{equation}
\left({x_1\over\sinh z}\right)^2+\left({x_2\over\sinh
  \delta}\right)^2=1.
\end{equation}

The relative sign of $z$ and $\delta$ determines the direction of the
orbit. In the special cases $\delta=\pm z$ the orbits become circular.
The timelike geodesic $\chi=0$ that appears to be at the centre of
each of these ellipses is not privileged physically. Rather, as the
background spacetime is maximally symmetric, any one timelike geodesic
can be transformed into $\chi=0$ by an isometry, and this will
transform all other timelike geodesics into ellipses about this new
centre.

Because all geodesics have the same period $2\pi$ in $t$ (and in
$\tau$), any two timelike geodesics that intersect once do so an
infinite number of times at coordinate time intervals $\Delta t=\pi$
and proper time intervals $\Delta\tau=\pi$, irrespective of their
relative boost at the points of intersection. This periodicity also
relates the timelike geodesics of padS3 and eadS3.


\subsection{Null geodesics}


To find null geodesics as a limit of our timelike geodesics, we take
the boost $z$ to infinity while also rescaling the proper time $\tau$
by an infinite factor. Reparameterising $\tau$ as
\begin{equation}
\label{lambdadef}
\lambda := \tau \cosh z
\end{equation}
and taking the limit $z\to\infty$, we obtain
\begin{subequations}
\begin{eqnarray}
\label{chitznull}
\chi(\lambda)&=&\sinh^{-1}\sqrt{\lambda^2+\sinh^2\delta},\\
t(\lambda)&=&\tan^{-1}\left({\lambda\over\cosh\delta}\right)+t_0, \\
\label{phitznull}
\phi(\lambda)&=&\tan^{-1}\left({\lambda\over\sinh\delta}\right)+\phi_0.
\end{eqnarray}
\end{subequations}

Null geodesics can of course also be found directly from
(\ref{timelikegeo}) with the right-hand side set to $0$ instead of
$-1$. The affine parameter $\lambda$ is defined only up to an affine
transformation. With (\ref{lambdadef}) we have normalised $\lambda$ so
that $dt/d\lambda=1/\cosh\delta$ at $\lambda=0$, which is the moment
of closest approach to $\chi=0$. In contrast to the timelike
case, for $-\infty<\lambda<\infty$, each null geodesic crosses eadS3
exactly once, entering and leaving through the timelike infinity at
$t=t_0\mp\pi/2$, $\chi=\infty$, $\phi=\phi_0\mp \pi/2$.


\subsection{Massive non-spinning particles}


An elegant way of computing isometry generators that leave a
given geodesic invariant was given in \cite{HM99}. This can be used to
find the generators of any nonspinning point particle. It is easily
verified that if we take any two points $\y$ and $\z$ and define
$u=\z\y^{-1}$ and $v=\y^{-1}\z$, then $\x=u^{-1}\x v$ is obeyed both
for $\x=\y$ and for $\x=\z$. But locally any two points $\y$ and $\z$
define a unique geodesic $\x(\tau)$ through them. Finally, it can be
shown that the fixed points of an isometry of padS3, if there are any,
must form a geodesic \cite{HM99}. Combining these two facts, we must
have that $\x(\tau)=u^{-1}\x(\tau)v$ for all points on the
geodesic. Hence $u$ and $v$ obtained from any two points
$\y=\x(\tau_\y)$ and $\z=\x(\tau_\z)$ on a geodesic are the generators of an
isometry that leaves precisely this geodesic invariant. It is clear
that the generator can depend only on $\nu:=\tau_\z-\tau_\y$.

The general timelike geodesic (\ref{chitz}-\ref{phitz}) in $SL(2,\R)$
notation, using (\ref{xtomat}) and (\ref{adS3parameterisation}), is
\begin{eqnarray}
\label{xdeltaz}
\x(\tau)&=&(\cos t_0\cosh\delta\cos\tau-\sin t_0\cosh z\sin\tau)\,I \nonumber \\
&+&(\cos t_0\cosh z\sin\tau+\sin t_0\cosh\delta\cos\tau)\,\gamma_0 \nonumber \\
&+&(\cos\phi_0\sinh\delta\cos\tau+\sin\phi_0\sinh z\sin\tau)\,\gamma_1\nonumber \\
&+&(\cos\phi_0\sinh
z\sin\tau-\sin\phi_0\sinh\delta\cos\tau)\,\gamma_2,\nonumber \\
\end{eqnarray}
where $\tau$ is proper time. The generators in terms of $\nu$,
$\delta$, $z$, $t_0$ and $\phi_0$ become
\begin{eqnarray}
\label{uvzdelta}
\left.\begin{aligned}u\\v\end{aligned}\right\}
&=&\cos\nu\,I 
+\sin\nu\Bigl\{\cosh z_\pm \,\gamma_0 \nonumber \\
&&+\sinh z_\pm 
\left[-\sin \phi_\pm\,\gamma_1+\cos \phi_\pm\,\gamma_2\right]\Bigr\},
\label{uvdeltaz}
\end{eqnarray}
where we have defined the shorthands
\begin{equation}
\label{phipmdef}
z_\pm:=z\pm\delta, \qquad \phi_\pm:=\phi_0\pm t_0.
\end{equation}
They obey the identities
\begin{equation}
u(t_0,\phi_0,z,\delta,\nu)=v(-t_0,\phi_0,z,-\delta,\nu)
\end{equation}
relating the left and right generator, and the rotation symmetry
\begin{eqnarray}
\label{rotationidentity}
u(t_0,\phi_0+\pi,z,\delta,\nu)=u(t_0,\phi_0,-z,-\delta,\nu)
\end{eqnarray}
and similarly for $v$. The latter symmetry is intuitive: reversing
both boost and impact parameter is equivalent to a rotation by $\pi$.
Note also that the geodesic in $SL(2,\R)$ notation and the particle
generators are regular as $\delta\to 0$, whereas the expressions for
$t$, $\chi$ and $\phi$ are not, due to the coordinate singularity of
the polar coordinates $(\chi,\phi)$ at $\chi=0$.

The generators for a particle 
sitting still at $\chi=0$, with $\delta=z=0$, are simply
\begin{equation}
\label{particleatrest}
u=v=\cos\nu\, I+\sin\nu\, \gamma_0,
\end{equation}
which agrees with the BTZ generators (\ref{BTZparticlegenerators}) for
zero spin. We read off
\begin{equation}
Tu=Tv=\cos \nu,
\end{equation}
and as the trace is invariant under conjugation, this expression is
invariant under the isometries that map one timelike geodesic to
another, and so it must be related to the rest mass of any nonspinning
particle, independently of location and velocity.  If we combine two
massive nonspinning particles on the same geodesic (characterized by
$\delta$, $z$, and $\phi_0$) by multiplying their generators, we find
that $(u_1u_2,v_1v_2)$ describe another particle on this geodesic,
with mass $\nu_1+\nu_2$. So $\nu$ is proportional to the locally
measured particle mass. The factor of proportionality is obtained by
directly solving the Einstein equations with the distributional
stress-energy tensor
\begin{equation}
T_{ab}=mu^au^b\delta(\x),
\end{equation}
which makes sense in 2+1 dimensions (only). The result is $\nu=4\pi m$
\cite{DeserJackiwTHooft84}, see also
Appendix~\ref{appendix:Lambda0}. Because the source is infinitesimally
small, this is independent of $\Lambda$.


\subsection{Massless non-spinning particles}


To obtain the massless limit of the massive particle generators, we
let $z\to\infty$ at the same time as $\nu\to 0$ such that 
\begin{equation}
\label{Wlimit}
\lim_{\nu\to 0}\nu\cosh z=:{W}
\end{equation}
is finite. We obtain
\begin{equation}
\left.\begin{aligned}u\\v\end{aligned}\right\}
=I +{W}e^{\mp\delta}\left[\gamma_0 -\sin \phi_\pm\,\gamma_1
+\cos \phi_\pm\,\gamma_2\right]. 
\label{uvmassless}
\end{equation}

Equivalently, we can repeat the construction of the particle
generators from the geodesic it is on. The null geodesics
(\ref{chitznull}-\ref{phitznull}) in $SL(2,\R)$ notation are
\begin{eqnarray}
\label{xlambdanull}
\x(\lambda)&=&(\cos t_0\cosh\delta+\lambda \sin t_0)\,I \nonumber \\
&+&(\sin t_0\cosh\delta+\lambda\cos t_0)\,\gamma_0 \nonumber \\
&+&(\cos\phi_0\sinh\delta -\lambda \sin\phi_0)\gamma_1 \nonumber \\
&+&(\sin\phi_0\sinh\delta +\lambda \cos\phi_0)\gamma_2.
\end{eqnarray}
We find (\ref{uvmassless}) again, with ${W}:=\lambda_\z-\lambda_\y$. Putting
two massless particles with energies ${W}_1$ and ${W}_2$ on the same
geodesic, the product of their generators has $W={W}_1+{W}_2$, so $W$ is
proportional to the particle energy, see also the end of
Sec.~\ref{section:twomasslessparticles} below for the relation between
$W$ and energy. 


\subsection{Massive spinning particle}


The approach starting from a geodesic cannot be extended to spinning
particles, as their symmetry identifies any point on the particle
world line with another one, shifted in time. However, we can obtain
the generators for a massive spinning particle by boosting and
displacing the BTZ solution. The 5-parameter family of isometries
(\ref{ghtransformation}) with generators
\begin{equation}
\label{ghdefs}
\left.\begin{aligned} g\\ h \end{aligned}\right\}
:=g_\text{gen}\left({z_\mp\over 2},\psi,\phi_\pm+\psi\right),
\end{equation}
where $g_\text{gen}$ was defined in (\ref{g0def}) and
$\phi_\pm$ was defined in (\ref{phipmdef}), is the most general one
(out of the 6-dimensional group of all isometries) which maps the
timelike ``central'' geodesic $\chi=0$ of the padS3 metric (\ref{adS3metric}),
represented in $SL(2,\R)$ notation as
\begin{equation}
\x_c=\cos(t_0+\tau)\,I+\sin(t_0+\tau)\,\gamma_0,
\end{equation}
to the geodesic (\ref{chitz}-\ref{phitz}), represented as
(\ref{xdeltaz}). We have already parameterised it so that it maps the
generators (\ref{BTZparticlegenerators}) of the nonspinning BTZ point
particle to (\ref{uvzdelta}) via the action
(\ref{transformedgenerators}).  

This 5-parameter family is periodic with period $2\pi$ in $\phi_\pm$
and $\psi$. Moreover, shifting $\psi$ by $\pi$ changes only the
overall sign of $g$ and $h$, and so corresponds to the same
isometry. Changing the signs of both $z$ and $\delta$ is equivalent to
shifting either $t_0$ or $\phi_0$ by $\pi$, as was the case for
non-spinning particles. Both $g$ and $h$ have a left factor
$\cos\psi\,I+\sin\psi\,\gamma_0$ that corresponds to a rotation by
$2\psi$ in the $x_1x_2$ plane (which leaves the particle trajectory
unchanged), applied before any boost and displacement. 

Hence we define the generators of a boosted and displaced spinning
point particle by (\ref{transformedgenerators}) with (\ref{ghdefs}), obtaining
\begin{eqnarray}
\label{spinninggenerators}
\left.\begin{aligned}u\\v\end{aligned}\right\}
&=&\cos\nu_\mp\,I+\sin\nu_\mp\,
\Bigl\{\cosh z_\mp \,\gamma_0 \nonumber \\
&&+\sinh z_\mp \,
\left[-\sin\phi_\pm\,\gamma_1+\cos\phi_\pm\,\gamma_2\right]
\Bigr\}. \nonumber \\
\end{eqnarray}
The value of $\psi$ does not affect the generators. As expected, these
generators map the geodesic (\ref{xdeltaz}) to itself but with a time
jump $\tau\to\tau+2\pi a_-$, compare (\ref{Deltathatdef}). In the
nonspinning case, we have $\nu_+=\nu_-=\nu$, and and
(\ref{spinninggenerators}) reduces to (\ref{uvzdelta}).


\subsection{Massless spinning particle}


We reparameterise
\begin{equation}
M=-1+{2 W\over \pi\cosh z},\quad J={2 U\over \pi\cosh z}
\end{equation}
The condition $|J|<M+1$ for the particle to be representable as an
isometry of padS3 is equivalent to $0<W<|U|$ (energy dominates spin). 

We now define the massless limit of (\ref{spinninggenerators}) by
letting $z\to \infty$ while keeping $W$, $U$ and $\delta$ fixed. The
resulting generators are
\begin{eqnarray}
\label{uvmasslessspinning}
\left.\begin{aligned}u\\v
\end{aligned}\right\}
&=&I+e^{\mp\delta}(W\mp U)\bigl[\gamma_0 \nonumber \\ &&
-\sin\phi_\pm\gamma_1+\cos\phi_\pm\gamma_2\bigr].
\end{eqnarray}
We recover the nonspinning case (\ref {uvmassless}) by setting $U=0$. 
Both $W$ and $U$ are additive if we place two massless particles on
the same orbit, so are proportional to the particle energy and spin.

The generator traces for a massless spinning particle are $Tu=Tv=1$,
as for the adS3 solution itself, so in contrast to the massive
spinning particles these are not BTZ solutions. It is clear that they
cannot be, as the BTZ solutions are by ansatz axisymmetric and
stationary, while the massless particle solutions are neither: the
particle crosses eadS3 at a specific time and in a specific direction,
which breaks these symmetries. 

The fact that only the generator traces are gauge-invariant, but they
are both 1, also suggests a single massless particle cannot be
distinguished from vacuum adS3 in a gauge-indepenent way. We can make
a gauge transformation from (\ref{uvmasslessspinning}) to
\begin{equation}
\left.\begin{aligned}\tilde u\\ \tilde v
\end{aligned}\right\}
=I+e^{\alpha_\mp}(W\mp U)(\gamma_0+\gamma_2),
\end{equation}
for two arbitrary (real and finite) gauge parameters $\alpha_\mp$. Hence not
only the orbital parameters $\delta$, $\phi_0$ and $t_0$ are
pure gauge, as one would expect, but so are 
$W\pm U$, up to the fact they do not vanish and their sign. 

What is the identification isometry parameterised by
(\ref{uvmasslessspinning})? The 1-parameter family of null geodesics
(\ref{xlambdanull}), with $\phi_0$ and $t_0$ fixed,
$-\infty<\delta<\infty$ labelling the geodesics, and $\lambda$ the
affine parameter along thm, form a null plane. This is most easily
seen in $\R^{(2,2)}$, where the family with, for simplicity,
$\phi_0=t_0=0$ is given by
\begin{equation}
x^\mu=(\cosh\delta,\lambda,\sinh\delta,\lambda).
\end{equation}
One can show that the isometry (\ref{uvmasslessspinning}) acts on the
null plane parameterised by $t_0$, $\phi_0$, $-\infty<\delta<\infty$
and $-\infty<\lambda <\infty$, where $\delta=\delta_0$ is the particle
worldline, as the identification
\begin{equation}
\lambda\sim \lambda +
2W\cosh(\delta-\delta_0)+2U\sinh(\delta-\delta_0),
\end{equation}
On the particle worldline itself, the shift in $\lambda$ is simply
$\Delta \lambda=2W$. Note it is neither an even nor an odd function of
$\delta-\delta_0$, and depends on both $W$ and $U$.


\section{Two spinning point particles}
\label{section:twoparticles}


 
\subsection{Interpretation of product order in the rest frame}


We now calculate the products of the generators of two point
particles. Interpreting this product physically requires some
care. Consider first two non-spinning massive particles with the same
mass $\nu_0$ and equal and opposite boosts $z_0$, colliding
head-on. Without loss of generality, we can make the particles collide
at $\chi=0$ at $t=0$, coming from the $y$ and $-y$ directions. Hence
it is sufficient to consider
\begin{equation}
\delta_i=t_{0i}=\phi_{01}=0, \quad \phi_{02}=\pi,
\quad \nu_i=\nu_0, \quad z_i=z_0\ge 0.
\end{equation}
The product generators in this special case are
\begin{eqnarray}
\label{headonproduct}
v_1v_2=u_1u_2&=&\left(\cos 2\nu \cosh^2 z_0-\sinh^2 z_0\right)I \nonumber \\
&&+\sin2\nu_0\cosh z_0\,\gamma_0\nonumber \\
&+&\sin^2\nu_0\sinh2 z_0\,\gamma_1.
\end{eqnarray}
They can be written as 
\begin{eqnarray}
v_1v_2&=&u_1u_2=\cos\nu_\text{tot}\,I \nonumber\\
&&+\sin\nu_\text{tot}(\cosh z_\text{tot}\,\gamma_0+\sinh z_\text{tot}\,\gamma_1)
\end{eqnarray}
with parameters 
\begin{subequations}
\begin{eqnarray}
\label{nuf}
\nu_\text{tot}&=&2\sin^{-1}(\sin\nu_0\cosh z_0), \\
\label{zf}
z_\text{tot}&=&-\tanh^{-1}(\tan\nu_0\sinh z_0).
\end{eqnarray}
\end{subequations}
These are the generators of a single particle with rest mass $\nu_\text{tot}$,
going through $\chi=0$ at $t=0$ but not sitting still there: rather,
the particle moves in the $x$-direction $\phi_0=\pi/2$ with a rapidity
$z_\text{tot}$. The appearance of this sideways boost is initially
surprising, but see Appendix~\ref{appendix:effectiveparticle} for the
visualisation of an example with $\Lambda=0$. As explained there, the
boost is reversed when the product order is reversed, that is
\begin{equation}
u_1u_2\leftrightarrow u_2u_1 \Leftrightarrow z_\text{tot}\leftrightarrow -z_\text{tot}
\end{equation}
and similarly for $v_2v_1$. This corresponds to two equally natural
ways of defining a coordinate system on the effective particle
spacetime. 

To put the joint particle at rest as seen from either one of the two
sides, we could apply a boost transformation to $u_1u_2$ with
generators
\begin{equation}
\label{boostgh}
g=h=g_\text{gen}\left(-{z_\text{tot}\over 2},0,{\pi\over 2}\right),
\end{equation}
where $g_\text{gen}$ was defined in (\ref{g0def}).
The result is 
\begin{equation}
g^{-1}u_1u_2g=h^{-1}v_1v_2h=\cos\nu_\text{tot}\,I+\sin\nu_\text{tot}\,\gamma_0,
\end{equation}
as intended. Alternatively, we could apply the opposite boost to the product
generators taken in the opposite order. 

 
\subsection{Rest frame and center of mass conditions}
\label{section:restframecm}


If there are only two massive, nonspinning particles in the universe,
the only physical parameters of the initial data are the rest masses
$\nu_i$ of the two particles and their relative rapidity $Z$ and
impact parameter $D$, measured in the ``rest frame'' of the system.

However, it is not obvious how to define the 3-momentum of a
self-gravitating test particle locally, as it is sitting at a
singularity of the metric. With $\Lambda<0$, the addition of the
3-momenta of two particles at different points also becomes ambiguous
as there is no parallelism at a distance. To give a precise definition
of the rest frame we need to define the center of mass of the system
at the same time.

We define a ``frame'' to be a patch of eadS3 that touches the
trajectories of both particles, together with a time slicing $t$ in
which each slice has the geometry $d\chi^2+\sinh^2\chi\,d\phi^2$ of
${\mathbb H}^2$. We define the distance $d(t)$ of the particles in that frame
as the length of the spatial geodesic linking them. We assume this
geodesic lies inside the patch and the wedges representing the
particles do not intersect it. Without loss of generality, let the
closest approach happen at $t=0$, with $D:=d(0)$.

We define as a necessary condition for the frame to be the rest frame
that the 2-velocities of two non-spinning particles are antiparallel
at $t=0$, where we compare them by parallel transport along the
spatial geodesic linking them. A sufficiently general family of
particles data is
\begin{equation}
\label{preCMassumptions}
t_{0i}=0, \quad\phi_{01}=0, \quad \phi_{02}=\pi, \quad z_i> 0.
\end{equation}
Clearly one of the periodic moments of closest approach, with respect
to the $t$-frame, is at $t=0$, and at that moment 
the particles move
in the $y$ and $-y$ directions, antiparallel by our operational
definition. We define the relative rapidity and impact parameter
(assumed to be measured in the rest frame) as
\begin{equation}
  Z:=z_1+z_2, \quad D:=\delta_1+\delta_2. 
\end{equation}
(Recall that rapidities $\gamma$, but not velocities $v$, are additive
in special relativity, and that the two are related by $v=\tanh
\gamma$.)  If we further define the center of mass to be at $\chi=0$,
both $\delta_i$ must have the same sign, but they, and hence $D$, can
have either sign. Reversing the sign of $D$ corresponds to the mirror
image of the initial data, and so reverses the orbital angular
momentum. In the rest frame, both $z_i$ must have the same
sign. Reversing the signs of $z_i,\delta_i$ corresponds to a rotation
of the whole system by $\pi$, which is pure gauge.

Assume initially that the two particles (\ref{preCMassumptions}) have
the same rest mass, offset and rapidity,
\begin{equation}
\label{symmetricassumptions}
\quad \nu_i=\nu, \quad \delta_i=\delta, \quad z_i=z.
\end{equation}
By symmetry, the effective particle, a spinning one, should be at rest
in the $t$-frame at the point $\chi=0$. The product generators under
the assumptions (\ref{preCMassumptions}) and
(\ref{symmetricassumptions}) are
\begin{eqnarray}
\left.\begin{aligned}u_1u_2\\v_1v_2\end{aligned}\right\} 
&=&\left(\cos 2\nu \cosh^2 z_\mp-\sinh^2 z_\mp\right)I \nonumber \\
&&+\sin2\nu\cosh z_\mp\,\gamma_0\nonumber \\
&&+\sin^2\nu\sinh2 z_\mp\,\gamma_1.
\end{eqnarray}
Reversing the product order in these expressions corresponds to
reversing the signs of both $\delta$ and $z$, which also reverses the
sign of the sideways boost of the effective particle (here, the sign
of the coefficient of $\gamma_1$). Geometrically, this represents a
rotation by $\pi$ of the original two-particle system.

Intuitively, this rotation symmetry should hold also for any
non-symmetric collision where the $z_i$ are measured in the rest frame
and the $\delta_i$ are measured relative to the center of mass.
Returning to the case where we assume only (\ref{preCMassumptions}),
but leave the $\nu_i$, $\delta_i$ and $z_i$ arbitrary, we therefore
{\em define} the center of mass and the rest frame by its symmetry
\begin{equation}
\label{rotationsymmetry}
(u_1u_2\leftrightarrow u_2u_1,v_1v_2\leftrightarrow v_2v_1)
  \Leftrightarrow
  (\delta_i\leftrightarrow-\delta_i,z_i\leftrightarrow-z_i).
\end{equation}
The $u_i$ and $v_i$ are given by (\ref{spinninggenerators}). Explicit
but tedious calculation shows that 
(\ref{rotationsymmetry}) holds if and only if the two constraints
\begin{equation}
\label{myuvcond}
\sinh z_{\pm 1} \tan\nu_1=(1\leftrightarrow2)
\end{equation}
on the initial data parameters hold, or equivalently 
\begin{subequations}
\label{restframecmconditions}
\begin{eqnarray}
\label{restframecondition}
\cosh\delta_1\sinh z_1\tan\nu_1&=&(1\leftrightarrow2), \\
\label{centerofmasscondition}
\sinh\delta_1\cosh z_1\tan\nu_1&=&(1\leftrightarrow2).
\end{eqnarray}
\end{subequations}
In the limit $\nu_i\ll 1$, $z_i\ll 1$ and $\delta_i\ll 1$ of a small, slow
orbit and small masses, where $\Lambda$, special-relativistic effects
and self-gravity can be neglected, these conditions become
\begin{subequations}
\begin{eqnarray}
\nu_1z_1&=&\nu_2z_2,  \\ 
\nu_1\delta_1&=&\nu_2\delta_2.
\end{eqnarray}
\end{subequations}
This limit identifies (\ref{restframecondition}) as the condition that the
$t$-frame is the rest frame and (\ref{centerofmasscondition}) as the
condition that $\chi=0$ is the center of mass. For head-on collisions,
with $\delta_i=0$, the center of mass condition duly becomes trivial,
and the rest frame condition becomes
\begin{equation}
\sinh z_1\tan\nu_1=(1\leftrightarrow2),
\end{equation}
in agreement with Eq.~(5.1) of \cite{L16b}.

For spinning particles, the rest frame and centre of mass conditions
(\ref{myuvcond}) generalise to
\begin{equation}
\label{myuvcondspinning}
\sin z_{\pm 1} \tan\nu_{\pm 1}=(1\leftrightarrow2),
\end{equation}
which must hold separately for the upper and lower signs. They do not
separate into center of mass and rest frame conditions of the form
(\ref{restframecmconditions}).

With
(\ref{preCMassumptions},\ref{restframecondition},\ref{centerofmasscondition})
imposed, the products of two generators (\ref{spinninggenerators}) are
\begin{eqnarray}
\label{prodgenmassrestframe}
\left.\begin{aligned}u_1 u_2\\v_1 v_2\end{aligned}\right\} 
&=&\bigl[\cos\nu_{\mp1}\cos\nu_{\mp2} \nonumber \\
&&-\sin\nu_{\mp1}\sin\nu_{\mp2}\cosh(Z\mp D)\bigr]\,I
  \nonumber \\ 
&&+\Bigl\{{1\over 2}\bigl[1-\cos2\nu_{\mp1}\cos2\nu_{\mp2} \nonumber \\
&&\qquad +\sin2\nu_{\mp1}\sin2\nu_{\mp2}\cosh(Z\mp D)\bigr]\Bigr\}^{1\over 2}
  \,\gamma_0 \nonumber \\ 
&&+\,\sin\nu_{\mp1}\sin\nu_{\mp2}\sinh(Z\mp D)\gamma_1.
\end{eqnarray}
By construction, the reverse product order is obtained by reversing
the signs of $Z$ and $D$. The trace of this expression takes this form
already in general gauge.

 
\subsection{Orbital parameters}


Solving $t(\tau_1)=t(\tau_2)$ for $\tau_2$ in terms of $\tau_1$ [with
  $t(\tau_i)$ given in (\ref{ttlz})], and using the rest frame and
centre of mass conditions (\ref{restframecmconditions}), we find that
$\phi_2(t)=\phi_1(t)+\pi$ for all $t$. In other words, in the center
of mass frame the two particles are always linked by a spatial
geodesic throught the point $\chi=0$, and so they appear to be
circling this point as one would intuitively expect for a common
center of mass in the rest frame. Of course, each particle actually
moves independently on an elliptic orbit. The center of mass and rest
frame conditions simply make these independent movements appear like
the effect of a mutual attraction, with force proportional to the
distance $r$.

We see from (\ref{chitz}) and (\ref{phitz}) that between $\tau=0$ and
$\tau=\pi/2$ the parameters $z_i$ and $\delta_i$ exchange roles, and
hence the same is true for their sums $D$ and $Z$. In scattering
theory language, $D$ is the impact parameter and $Z$ the relativity
rapidity. For periodic orbits, and with our convention that $|D|\le
Z$, $D$ is also the apogee distance, signed with the handedness of the
orbit, and $Z$ the perigee distance.

 
\section{The product state and its geometric interpretation}
\label{section:finalstate}


 
\subsection{Total mass and angular momentum of the two-body system}


We now come to the application of (\ref{MJfromTuTv}).  As the two-body
systems we consider here cannot radiate or divide their energy and
angular momentum, and as the generator traces are invariant under
isometries, $M$ and $J$ computed from a suitable product of the
generators of the two individual objects by using must be the total
mass and angular momentum. Applying (\ref{MJfromTuTv}) to the product
generators we have
\begin{subequations}
\label{MJfromTuTvbis}
\begin{eqnarray}
M_\text{tot}- J_\text{tot}&=&{\cal Q}(Tu_1u_2), \\
M_\text{tot}+J_\text{tot}&=&{\cal Q}(Tv_1v_2),
\end{eqnarray}
\end{subequations}
where the function ${\cal Q}(T)$ was defined in (\ref{calQdef}).  We
identify the final state as a black hole if $M>|J|$. This does not say
anything about the process of black hole formation, which we do not
examine here.

However, Holst and Matschull \cite{HM99} have constructed the full
spacetime for two massless non-spinning point particles, which enter
through the conformal boundary. They show that if and only if the
effective state is a black hole, a black hole is formed and the two
particles fall through its event horizon. Based on this work, we {\em
  conjecture} that when the effective state for any two point
particles (massive or massless, spinning or non-spinning) is a black
hole all particles fall into a black-hole horizon. When one or both of
the particles are massive, they emerge from a white-hole horizon. When
both particles are massless and enter through the conformal boundary,
the white hole is absent. To avoid the white hole we could create one
or both of the massive particles in a collision of massless particles
that themselves have entered through the conformal boundary of adS3.

We will show for two massive point particles that a point-particle
total state corresponds to a binary that orbits forever if
$\nu_1+\nu_2<\pi$, but a closed universe if $\nu_1+\nu_2>\pi$, always
with $0<\nu_i<\pi$. To complement this, we also {\em conjecture} that an
overspinning particle effective state corresponds to a binary system that
orbits forever, again based on the result of \cite{HM99} for two
massless non-spinning particles. 

 
\subsection{Two massive spinning particles}


Recall from (\ref{prodgenmassrestframe}) that the traces of the
product generators in the massive spinning case are
\begin{eqnarray}
\label{productmassivespinning}
\left.\begin{aligned}Tu_1u_2\\Tv_1v_2\end{aligned}\right\}
&=&\cos\nu_{\mp1}\,\cos\nu_{\mp2}\nonumber \\
&&-\sin\nu_{\mp1}\,\sin\nu_{\mp2}\,\cosh(Z\mp D).
\end{eqnarray}
Both are $\le 1$, so $M_\text{tot}$ and $J_\text{tot}$ are
defined from (\ref{MJfromTuTvbis}), and  obey $|J_\text{tot}|<-1+M_\text{tot}$.

\begin{figure}
\includegraphics[scale=0.6, angle=0]{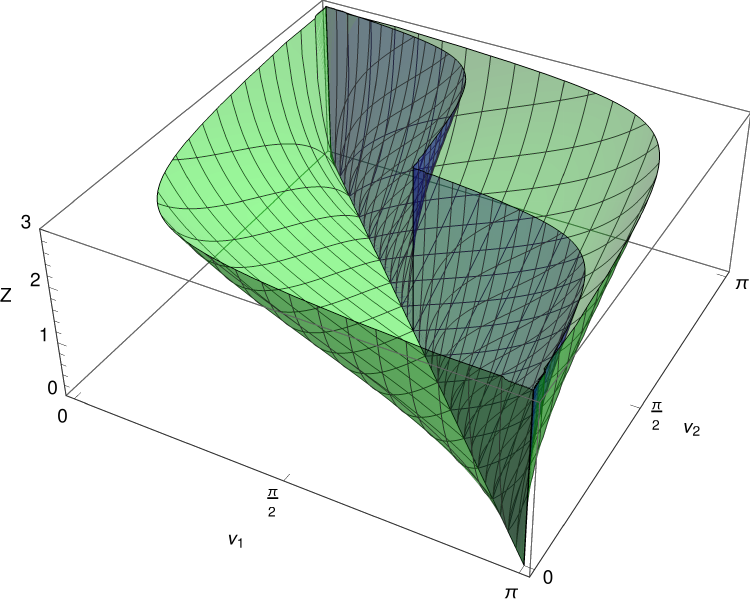} 
\caption{A plot of the possible effective states in the 3-dimensional
  parameter space of two nonspinning point particles that collide
  head-on, such that the solution has a moment of time symmetry: the
  parameters are $0\le\nu_1,\nu_2<\pi$ and $Z\ge 0$, which is equal
  both to the relative rapidity at impact, and the distance when the
  particles are momentarily at rest. (The particle spins and the
  impact parameter $D$ are assumed to be zero.) The boat-shaped green
  surface is $Z=C(\nu_1,\nu_2)$. Below and to the left of this surface
  (small $\nu_i$, small $Z$) the effective state is a virtual
  effective particle. Below and to the right (large $\nu_i$, small
  $Z$) the effective state is real third particle, which closes
  space. Above the surface (large $Z$), the effective state is a black
  hole. Within this black-hole region, for data to the right of the
  blue surface (large $\nu_i$) the moment of time symmetry (when the
  two particles are at maximum separation) already contains an
  apparent horizon.
\label{figure:Steif2particleplot}}
\end{figure}

From our convention that $Z\ge |D|$ we find that $|Z\pm D|=Z\pm
D$. Then from (\ref{productmassivespinning}) and (\ref{MJfromTuTvbis}) we have
\begin{subequations}
\label{massivethreshold}
\begin{eqnarray}
M_\text{tot}-J_\text{tot}>0 &\quad \Leftrightarrow\quad & Z-D>C_-, \\
M_\text{tot}+J_\text{tot}>0 &\quad \Leftrightarrow\quad & Z+D>C_+,
\end{eqnarray}
\end{subequations}
where we have defined the shorthands
\begin{equation}
\label{Cpmdef}
C_\pm := \cosh^{-1}\left({1+\cos\nu_{\pm 1}\cos\nu_{\pm 2}\over 
\sin\nu_{\pm 1}\sin\nu_{\pm 2}}\right),
\end{equation}
By the assumption that both individual objects are point particles,
with $|J_i|<-M_i$, the argument of $\cosh^{-1}$ in (\ref{Cpmdef}) is
$\ge 1$, and so $C_\pm$ defined in (\ref{Cpmdef}) are real. We choose
the positive branch of $\cosh^{-1}$, so that $C_\pm\ge 0$, with
equality at $\nu_{\pm 1}+\nu_{\pm 2}=\pi$.  In the nonspinning case we
have
\begin{equation}
\label{Cdef}
C_+=C_-=C:=\cosh^{-1}\left({1+\cos\nu_1\cos\nu_2\over\sin\nu_1\sin\nu_2}\right).
\end{equation}
The graph of the function $C(\nu_1,\nu_2)$ is the green surface in
Fig.~\ref{figure:Steif2particleplot}.

\begin{figure}
\includegraphics[scale=0.75, angle=0]{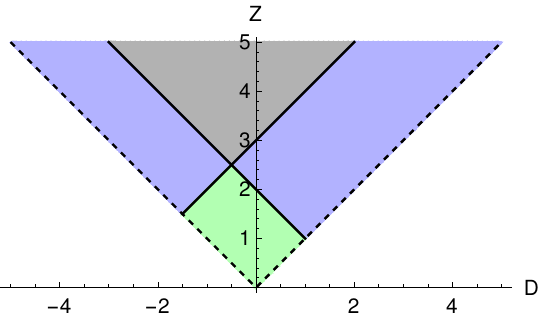} 
\caption{A plot of the different effective states for two massive
  spinning particles, plotted in the $(D,Z)$-plane, for arbitrarily
  fixed values $C_+=3$ and $C_-=2$. The plot continues to larger $Z$
  and $|D|$. By definition $Z\ge|D|$, for $D$ to be interpreted as the
  (signed) impact parameter and $Z$ the relativity rapidity of the two
  particles at the moment of closest approach, or $|Z|$ as the apogee
  and $D$ as the perigee. In the green region, for
  $0\le\nu_1+\nu_2<\pi$ spacetime at infinity is that of an effective
  spinning particle and the binary is eternal, while for
  $\nu_1+\nu_2>\pi$ space is closed by a real third particle and
  collapses. Blue corresponds to an overspinning particle, and grey to
  a spinning black hole. The regions and diagonal lines in the $(D,Z)$
  plane here correspond to the regions and lines of the same colour in
  Fig.~\ref{figure:JMplane}.
\label{figure:fig3}}
\end{figure}

Fig.~\ref{figure:fig3} shows the total state by colour-coding in the
$(D,Z)$-plane, for particle masses and spins $\nu_{\pm i}$ that give
rise to values $C_-=2$ and $C_+=3$, chosen arbitrarily for this
plot. The colour-coding is the same as in Fig.~\ref{figure:JMplane},
and we see that regions of parameter space representing black holes,
point particles and overspinning particles are laid out in
qualitatively the same way in the $(Z,D)$-plane of orbital parameters
as in the $(J,M)$ plane of BTZ states. The boundary $Z=|D|$ of the
plot is just our convention that $Z\ge|D|$. The lines $Z=C_\mp\pm D$
that separate black hole, point particle and overspinning particle
outcome regions depend on the particle masses and spins $\nu_{\pm i}$
through the two combinations $C_\pm$.

The 3-dimensional subcase of our 6-dimensional space of initial data
where two non-spinning massive particles collide head-on, and so the
spacetime admits a moment of time symmetry has previously been
investigated by Steif \cite{Steif96}. We have chosen a gauge where
$u=v$ if and only if there is a time-symmetry, and this single element
of $SL(2,\R)$ then parameterises the isometries of the 2-geometry of
the moment of time-symmetry. (See Appendix~\ref{appendix:Steif} for a
summary.) This special case already illustrates a subtlety: A
point-particle effective state can represent either a (fictitious)
effective point particle, and hence again a binary system that orbits
forever, or a genuine third particle that closes space.  In this
latter case we have three particles on an equal footing, two of which
we specified arbitrarily and a third that is determined by the first
two. All three particles emerge from a big-bang singularity and end in
a big-crunch singularity.

We now show that a point-particle effective state means an
effective particle, and so a binary system orbiting forever, if and only if 
$\nu_1+\nu_2<\pi$, and a real third particle and closed space if
and only if $\nu_1+\nu_2>\pi$, independently of the orbital parameters
and particle spins.

To see this, consider first the limit where $Z$ and $D$ are much
smaller than one. The two-particle orbit is then much smaller than the
cosmological length scale $\ell$ and so spacetime outside the
particles can be approximated as locally flat, and the particles are
moving slowly, so special-relativistic effects can also be
neglected. In this limit, the generator traces
(\ref{productmassivespinning}) are approximately
\begin{eqnarray}
\left.\begin{aligned}Tu_1u_2\\Tv_1v_2\end{aligned}\right\}
&\simeq &\cos(\nu_{\mp1}+\nu_{\mp 2}) \nonumber \\ 
&=& \cos\left(\nu_1+\nu_2\mp{\Delta {\hat t}_1+\Delta {\hat t}_2\over 2}\right) \nonumber \\ 
&=:&\cos\left(\nu_\text{tot}\mp {\Delta {\hat t}_\text{tot}\over2}\right).
\label{pointparticletracesZDzero}
\end{eqnarray}
Because of the periodicity of the cosine, this has multiple inverses
for $\nu_\text{tot}$ and $(\Delta \hat t)_\text{tot}$, and we now need to
consider which choice of inverse is physical.

If $0\le\nu_1+\nu_2\le\pi$, the total defect angle is less than
$2\pi$, and the excision wedges of the two particles (which in the
$\Lambda=0$ approximation are straight lines) can be rotated so that
they do not overlap. (Where one locates the wedges is otherwise pure
gauge.)  One then has a picture of time slices similar to that of
Fig.~\ref{figure:fig1} (in Appendix~\ref{appendix:Steif}): space is
open, and outside both particles the spatial geometry is that of a
cone, while seen from infinity the spatial slice is also a cone with
vertex at a fictitious effective particle. From elementary Euclidean
geometry, the defect angles add up to the total defect
angles. Similarly, the time jumps add up when we go around both
particle world lines.  Hence the correct solution of
(\ref{pointparticletracesZDzero}) is
\begin{equation}
\nu_\text{tot}\simeq \nu_1+\nu_2, \qquad 
\Delta\hat t_\text{tot}\simeq \Delta\hat t_1+\Delta\hat t_2.
\end{equation}
(We have written $\simeq$ as a reminder that this is only an
approximation for small, slow orbits.)

For $\nu_1+\nu_2>\pi$, the excision wedges have to overlap. Only a
four-sided compact region of each time slice is now physical. If the
particles are at A and B, and their excision wedges intersect at C and
C', for consistency we now {\em must} rotate the wedges so that C is
mapped to C' by the identification associated with either particle, in
other words AC has the same length as AC', and BC has the same length
as BC'. The wedges now face each other symmetrically, and the
triangles ABC and ABC' are equal up to a reflection. 

The vertices C and C' now both represent a third particle that closes
space: AC is glued to AC', and BC to BC', so that C and C'
coincide. Thus glued together, all three particles are on an equal
footing. However, in the opened-up form, the entire physical angle
around particle A is the angle CAC', and similarly the entire physical
angle around particle B is the angle CBC', but the physical angle
around particle C is split equally into the two sectors ACB and AC'B.

It is easy to see that for small orbits, where the curvature of
spacetime can be neglected, the sum of physical angles around
particles A, B and C is $2\pi$, namely the sum of interior angles of
the two triangles ABC and ABC'. This means that the sum of the {\em
  defect} angles at the three particles is $3\times
2\pi-2\pi=4\pi$. This is the solid angle of $S^2$: each time slice has
topology $S^2$, but the entire solid angle is concentrated at the
three particles, with space flat in between. Beyond the small-orbit
approximation, the curvature of space due to the cosmological constant
also becomes significant. As this is negative for $\Lambda<0$, the sum
of defect angles must be larger than $4\pi$.

With space closed, the total time jump going around all
particles must be zero, as any loop going round all three particle
world lines is contractible, that is, going around none. The correct
solution of (\ref{pointparticletracesZDzero}) for
$\pi<\nu_1+\nu_2<2\pi$ is therefore
\begin{equation}
\label{threeparticlesolution}
\nu_1+\nu_2+\nu_3\simeq 2\pi, \qquad 
\Delta\hat t_1+\Delta\hat t_2+\Delta\hat t_3\simeq0.
\end{equation}
Note that in this entire argument the wedge {\em surfaces} are
independent of the particle spins, which only affect the time shift
made in the identification, and so an effective particle is real if
and only if $\nu_1+\nu_2>\pi$, independently of the particle spins.
We now show that in the generic case, where $Z$ and $D$ are not
small, this simple algebraic criterion still holds exactly.

As we have seen, the particle is real if the forward excision wedges
overlap, and virtual if the backward excision wedges overlap. Suppose
we start from a situation where the effective particle is real, and
then reduce the defect angles of the two particles, while keeping
their orbits and time jumps fixed. The real particle will move to the
boundary of the adS3 cylinder, and return as a virtual particle. Hence
the transition occurs when the effective particle is at infinity. This
occurs when either $z$ or $\delta$ of the effective particle become
infinite (an elliptic orbit reaching infinity), or both (a circular
orbit at infinity). Hence, one or both of $\cosh z_\pm$ of the
effective particle become infinite at the transition from real to
effective particle.

Note now that when the matrices $u_i$ and $v_i$ are bounded, then so
are $u_1u_2$ and $v_1v_2$. From (\ref{spinninggenerators}) we can
write
\begin{equation}
\label{spinninggeneratorsbis}
\left.\begin{aligned}u_1u_2\\v_1v_2\end{aligned}\right\}
=\cos\nu_{{\rm tot}\mp}\,I+\sin\nu_{{\rm tot}\mp}\left(\cosh z_{{\rm tot}\mp}\,\gamma_0 +\dots\right),
\end{equation}
and this must remain bounded as the effective particle reaches
infinity.  Given that the coefficients of $\gamma_0$ in
(\ref{spinninggeneratorsbis}) must remain finite but one or both of
$\cosh z_{{\rm tot}\mp}$ become infinite, we must have that the
corresponding $\sin\nu_{{\rm tot}\mp}=0$ and so the corresponding
$\cos\nu_{{\rm tot}\mp}=1$ or $-1$. In other words, an effective
particle can be on an orbit reaching infinity only if either one of
$Tu_1u_2$ or $Tv_1v_2$ is equal to either $1$ or $-1$, or both are.

Looking at the product traces now as the functions
(\ref{productmassivespinning}) of the initial parameters we see that,
for fixed $\nu_{i\pm}$, $T=\pm1$ occurs at the boundaries of the
point-particle region in the $(D,Z)$ plane, so the effective particle
cannot change nature within that region. But then we can choose to
evaluate its nature in the small, slow orbit approximation $|D|,Z\ll 1$ as
above.  In other words, the effective particle is virtual for
$\nu_1+\nu_2<\pi$ and real for $\nu_1+\nu_2>\pi$, independently of the
$\Delta\hat t_i$, $D$ and $Z$.


\subsection{Two massless spinning particles}
\label{section:twomasslessparticles}


We now consider the case of two massless spinning particles. The
special case of two non-spinning massless particles was treated in
\cite{HM99}.  The traces of the product generators are
\begin{eqnarray}
\label{productmasslessspinning}
\left.\begin{aligned}Tu_1u_2\\Tv_1v_2\end{aligned}\right\}
&=&1-2e^{\mp D}(W_1\mp U_1)(W_2\mp U_2).
\end{eqnarray}

In the massless limit, the rest frame and centre of mass conditions
(\ref{myuvcondspinning}) become
\begin{equation}
\label{myuvcondspinningbothmassless}
e^{\pm \delta_1}(W_1\pm U_1)=(1\leftrightarrow2).
\end{equation}
These two equations can be rearranged as 
\begin{equation}
e^{\delta_1-\delta_2}={W_2+U_2\over W_1+U_1}={W_1-U_1\over W_2-U_2}.
\end{equation}
and so $W_1$, $W_2$, $U_1$ and $U_2$ measured in the rest frame must
obey the one constraint
\begin{equation}
\label{masslessspiningconstraint}
W_1^2-U_1^2=W_2^2-U_2^2.
\end{equation}
As we have $W_i>|U_i|$, we can parameterise this constraint as
\begin{equation}
\label{sigmadef}
W_i=W_0\cosh\sigma_i, \quad U_i=W_0\sinh\sigma_i
 \end{equation}
where $W_0>0$ and $-\infty<\sigma_i<\infty$ can now be chosen
freely. The two constraints (\ref{myuvcondspinningbothmassless}) then
both reduce to
\begin{equation}
\delta_1+\sigma_1=\delta_2+\sigma_2.
\end{equation}
Given the gauge-invariant parameters $D$, $\sigma_1$ and $\sigma_2$,
this can be solved for
\begin{equation}
\label{Tproductbothmasslessspinning}
\left.\begin{aligned}\delta_1\\ \delta_2\end{aligned}\right\}={D\pm(\sigma_2-\sigma_1)\over
    2}.
\end{equation}
We then find
\begin{equation}
\label{uvmasslessspinningbis}
\left.\begin{aligned}Tu_1u_2\\Tv_1v_2\end{aligned}\right\}
=1-2W_0^2 e^{\mp (D+\Sigma)},
\end{equation}
where 
\begin{equation}
\label{Sigmadef}
\Sigma:=\sigma_1+\sigma_2,
\end{equation}
and the four gauge-invariant parameters $W_0$, $\sigma_1$, $\sigma_2$
and $D$ can be chosen freely. Substituting (\ref{Tproductbothmasslessspinning})
into (\ref{MJfromTuTvbis}) gives
\begin{equation}
M\pm J={\cal Q}\left(1-2W_0^2 e^{\pm(D+\Sigma)}\right),
\end{equation}
and so $J$ is an odd analytic function of $D+\Sigma$ while $M$ is
even.

The inequalities for classifying the effective state are
\begin{subequations}
\begin{eqnarray}
M_\text{tot}-J_\text{tot}>0 &&\quad \Leftrightarrow\quad D+\Sigma<2\ln W_0, \\
M_\text{tot}+J_\text{tot}>0 &&\quad \Leftrightarrow\quad D+\Sigma>-2\ln W_0.
\end{eqnarray}
\end{subequations}
We see that it is a black hole state if and only if 
\begin{equation}
|D+\Sigma|<2\ln W_0,
\end{equation}
which obviously requires $W_0>1$, a point particle if and only 
\begin{equation}
|D+\Sigma|<-2\ln W_0,
\end{equation}
which requires $W_0<1$, and otherwise an overspinning particle
state. We have the intuitive result that it is, qualitativey, the sum
of orbital angular momentum and particle spins that impedes
collapse. The non-spinning case is recovered simply by setting
$\Sigma=0$, and for this case \cite{HM99} have shown that the black
hole effective state corresponds to dynamical black hole formation,
while for both the point particle and overspinning particle effective
state the two particles leave through the conformal boundary. We {\em
  conjecture} that this holds also in the spinning case.

In the special case where two counterspinning particles of the same
energy collide headon and merge into a single non-spinning massive
point particle, or nonspinning black hole, we have $D=\Sigma=0$, and
$\nu$ or $M$ are related to $W_0$ by $1-2W_0^2=\cos\nu$ for $W_0<1$ or
$2W_0^2-1=\cosh\sqrt{M}$ for $W_0\ge 1$.

 
\subsection{One massive and one massless spinning particle}


In the mixed case where particle~1 is massive and particle~2 is
massless, and reparameterising
\begin{equation}
\label{sigma2def}
W_2=:W_{02}\cosh\sigma_2, \quad U_2=:W_{02}\sinh\sigma_2,
\end{equation} 
the traces of the product generators are
\begin{equation}
\label{productmixedspinning}
\left.\begin{aligned}Tu_1u_2\\Tv_1v_2\end{aligned}\right\}
=\cos\nu_{\mp1} -\sin\nu_{\mp 1}\,W_{20}e^{z_1\mp (D+\sigma_2)}.
\end{equation}

The rest frame and centre of mass conditions (\ref{myuvcondspinning})
become
\begin{equation}
\label{myuvcondspinningmixed}
\sin(z_1\pm\delta_1) \tan\nu_{\pm 1}=W_{20}e^{\mp(\delta_2+\sigma_2)}.
\end{equation}
Writing $\delta_2=D-\delta_1$ and eliminating $\delta_1$,
we obtain
\begin{equation}
\label{z1quarticspinning}
K^2+2K\cosh\tilde D+1=e^{4z_1}
\end{equation}
where we have defined the shorthands
\begin{subequations}
\label{mixedspinningdefs}
\begin{eqnarray}
\label{Kdefspinning}
K&:=&{2e^{z_1}W_{20}\over\sqrt{\tan\nu_{+1}\,\tan\nu_{-1}}}, \\
\alpha_1 &:=& {1\over2}
\ln{\tan\nu_{-1}\over\tan\nu_{+1}}, \\
\tilde D&:=&D+\sigma_2+\alpha_1.
\end{eqnarray}
\end{subequations}
$\tilde D$ is an impact parameter corrected for the spin of the
massive particle (through $\alpha_1$) and the massless particle
(through $\sigma_2$).  We solve (\ref{z1quarticspinning}) as a
quadratic equation for $K$, obtaining
\begin{equation}
\label{Ktildesolution}
K=-\cosh \tilde D+\sqrt{e^{4z_1}+\sinh^2\tilde D}.
\end{equation}
Hence the product generator traces in their final form are
\begin{equation}
\label{Tproductmixedspinning}
\left.\begin{aligned}Tu_1u_2\\Tv_1v_2\end{aligned}\right\}
=\cos\nu_{\mp1}  -{\sin\nu_{\mp1}\,\tan\nu_{\mp1}\over 2} 
e^{\mp \tilde D}K,
\end{equation}
where the five gauge-invariant parameters $\nu_{\pm1}$, $z_1$, $D$ and
$\sigma_2$ can now be specified freely, and $K$ is given by
(\ref{Ktildesolution}).  In the nonspinning case $J_1=U_2=0$ we have
$\alpha_1=\sigma_2=0$ and $\tilde D=D$.

The inequalities for classifying the product isometry are
\begin{subequations}
\begin{eqnarray}
M_\text{tot}-J_\text{tot}>0 &\quad \Leftrightarrow\quad & \tilde D>\tilde C_-, \\
M_\text{tot}+J_\text{tot}>0 &\quad \Leftrightarrow\quad & \tilde D<\tilde C_+,
\end{eqnarray}
\end{subequations}
where we have defined the shorthands
\begin{eqnarray}
\tilde C_{\pm}&:=&\pm {1\over 2}\ln 
{e^{4z_1}-1-\hat C_\mp\over \hat C_\mp(\hat C_\mp+1)}, \\
\hat C_\pm&:=&2\,{1+\cos\nu_{\pm 1}
\over \sin\nu_{\pm 1}\,\tan\nu_{\mp1}}.
\end{eqnarray}
In particular, we the effective state is a black hole for
\begin{equation}
\tilde C_-<\tilde D<\tilde C_+
\end{equation}
(which implies $\tilde C_-<\tilde C_+$), an effective point particle
for
\begin{equation}
\tilde C_->\tilde D>\tilde C_+
\end{equation}
(which implies $\tilde C_->\tilde C_+$), and an effective overspinning
particle otherwise. We {\em conjecture} that a black hole forms
dynamically if and only if the effective state is a black hole, and
otherwise the massless particle leaves the spacetime through the
conformal boundary.


\section{Conclusions}
\label{section:conclusions}


We have taken the first steps in a research programme of classifying
all solutions of 2+1-dimensional general relativity with negative
cosmological constant that contain two compact objects surrounded by
vacuum. 

The programme has three key ingredients: First, the exterior of any
compact object must be an identification of adS3 under an
isometry. Second, the same must be true for the exterior of the
composite object. Third, we can obtain the composite isometry as a
product of the component isometries representing two objects and their
relative motion, as was first done in the $\Lambda=0$ case by Deser,
Jackiw and t'Hooft \cite{DeserJackiwTHooft84}.

The possible exterior solutions for massive compact objects are
precisely the BTZ axistationary vacuum solutions \cite{BTZ92}, which
are parameterised by any real values of mass $M$ and spin $J$. These
can be black holes $M\ge |J|$, point particles $M<-|J|$ and
overspinning particles $|J|>|M|$. The black-hole solutions are very
well studied, the particle solutions less so. In assembling our
building blocks, we have filled a few small gaps in the literature,
such as the maximal analytic extensions of the particle solutions and
the nature of their singularities, and the construction of the
overspinning particle solutions as identifications of adS3.

We can also admit the point particle and overspinning particle
solutions on the entire domain $0<r<\infty$ as particle-like
solutions. $r=0$ is a singular worldline only for non-spinning point
particles, but otherwise is the boundary of a region of closed
timelike curves that must be excised. (The singular worldline $\chi=0$ is
inside this region.) 

Finally, we can boost point particles while reducing their rest
mass in order to create perfectly sensible massless (spinning
particles) \cite{HM99}, and so these should be considered as
well. (This probably makes no sense for extended objects with point
particle exterior). This completes our list candidates for the two
objects. 

We have encountered two major difficulties in this programme. The
first one is that while all real values of $(J,M)$ correspond to
distinct BTZ solutions, the algebraic representation of the isometries
of adS3 as the group $SL(2,\R)\times SL(2,\R)\times/\Z$ \cite{BHTZ93}
covers only the segment $M>-1+|J|$ of the $(J,M)$ plane, comprising
all black hole solutions but not all point particle or overspinning
particle solutions.

The solutions we miss either have a time shift in their identification
that is larger than $2\pi$, and/or an excess, rather than deficit,
angle. There seems to be no reason for excluding these, and it is
possible that the wider isometry group that generates them can be
represented in a useful way. On the other hand, the space of rigidly
rotating perfect fluid stars with causal barotropic equations of state
constructed in \cite{2+1rotstars} fill precisely the region $M>-1+|J|$
of the $(J,M)$ plane, so it seems also possible that the exteriors of
physically reasonable compact objects all fall into this region, which
can be represented by the isometry group of padS3.

We can set aside this first difficulty by restricting ourselves to
objects with $M_i>-1+|J_i|$ for $i=1,2$. Within this category, we have
further restricted the two bodies to be to massive point particles or
compact objects with massive point particle exterior, which have
$-1+|J_i|<M_i<-|J_i|$ (the dark green region in
Fig.~\ref{figure:JMplane}), or massless point particles, which have
$M_i=-1$, $J_i=0$ (but non-trivial isometry generators). We have
computed $M_\text{tot}$ and $J_\text{tot}$ of the composite object as
explicit analytic functions of six gauge-invariant parameters: the two
rest masses and spins, and the impact parameter and energy in the
center of mass frame.

The second major difficulty is how to draw conclusions about the
global spacetime from the algebraic calculation. We have not attempted
this, but rely on Holst and Matschull's beautiful explicit
construction of the global spacetime with two massless nonspinning
point particles \cite{HM99}.  Based on their work, we {\em conjecture}
the following: if the product isometry corresponds to a black hole, a
black hole is actually created dynamically. When the composite state
of two massive particles is an overspinning particle, the binary is
eternal. When it corresponds to a massive point particle, the binary
is either eternal, or the effective particle is real and closes space,
which then recollapses. We have found a simple criterion for which of
the last two possibilities is realised. Finally, when either one or
both of the two bodies is a massless particle and no black hole is
formed the massless particle(s) leave the spacetime through the
conformal boundary. 

Our expressions for $M_\text{tot}$ and $J_\text{tot}$ remain finite and
analytic at the black hole threshold. In this sense, there is no
Choptuik-style critical scaling \cite{Choptuik92,GundlachLRR}. The
same is true also for toy models in 2+1 dimensions, such as dust balls
\cite{MannRoss93,VazKoehler08}, thin dust shells \cite{PelegSteif95},
or shells with tangential pressure \cite{MannOh06} and rotation
\cite{MannOhPark09}. All these systems cannot shed mass or spin so if
a black hole is formed it contains all the mass and spin, and so the
threshold of black hole formation is
$|J_\text{tot}|=M_\text{tot}$. (By contrast, for 3+1 spacetime
dimensions Kehle and Unger \cite{KehleUnger24} have made the exciting
conjecture that there are regions of solution space where the
threshold solutions are extremal black holes also for more physical
matter models or even vacuum.)

However, a system that {\em can} shed mass and spin during collapse
will show genuinely non-trivial critical phenoma even in 2+1 spacetime
dimensions. This has been demonstrated for an axisymmetric massless
scalar field without and with angular momentum in
\cite{JalmuznaGundlachChmaj2015,JalmuznaGundlach2017}, and for an
axisymmetric perfect fluid without and with angular momentum in 
\cite{BourgGundlach21a,BourgGundlach21b}.

What remains to be done to complete the solution of the two-body
problem is to investigate the spacetimes where the two bodies include
overspinning particles and black holes, or compact objects with such
exteriors. We have not attempted this as we are not sure about the
geometric meaning of the traces of the generator products. However,
given that there are no periodic test particle orbits on black hole
spacetimes, it is unlikely that eternal black hole binaries
exist. Rather, it is likely that an effective point particle or
overspinning particle is real and closes space, as in the solutions
found in \cite{Clement94}.

It would also be interesting to see if BTZ solutions with $M>-1+|J|$
can be realised as identifications under an isometry of a spacetime we
have called uadS3, with these isometries parameterised in form that
allows them to be composed explicitly. 

\acknowledgments

The author is grateful to Gavin Hartnett, Konstantinos Skenderis and
EPSRC-funded PhD student Andrew Iannetta for stimulating conversations
while this work was begun. Adrien Loty contributed to the conformal
compactifications and the distinction between two-particle and
three-particle solutions in an internship funded by \'Ecole
Polytechnique. This paper would not have been completed without the
patient explanations and probing questions of Jorma Louko. All errors are of
course my own.


\begin{appendix}



\section{$SO(2,2)$ and $SL(2,\R)\times SL(2,\R)/{\mathbb Z}^2$}
\label{appendix:SO22SL2R}


It is helpful to translate the pair of $SL(2,\R)$ generators into a
single $SO(2,2)$ generator to see how the isometry acts on the
hyperboloid (\ref{hyperboloid}) embedded in $\R^{(2,2)}$. We define the
matrix $R\in SO(2,2)$ equivalent to the pair $(u,v)\in SL(2,\R)\times
SL(2,\R)/{\mathbb Z}_1$ by
\begin{equation}
(RX)\cdot\gamma:=u^{-1}(X\cdot\gamma)v,
\end{equation}
for all $X\in \R^{(2,2)}$, where $X\cdot\gamma$ is shorthand for
(\ref{xtomat}). Recall that we use the coordinates
$X=(x_3,x_0,x_1,x_2)$, where we denote the two timelike directions by
$(x_3,x_0)$. This definition gives the explicit formula
\begin{equation}
\label{Rformula}
R_{\mu\nu}={1\over 2}(-1)^{\delta_{\mu 0}}\tr\left(
\gamma_\mu u^{-1}\gamma_\nu v\right)
\end{equation}
for the components of the matrix $R$.
If we write $u=\sum u_\mu\gamma_\mu$, where
$u_0^2+u_3^2-u_1^2-u_2^2=1$, and similarly for $v$, then each matrix element
$R_{\mu\nu}$ of $R$ is a sum of four terms of the form $u_\alpha
v_\beta$. 

Consider now the $SO(2,2)$ matrix $S$ equivalent to the pair of
$SL(2,\R)$ matrices $(u,I)$, and $T$ equivalent to the pair
$(I,v)$. The generators of the product of these isometries, in either
order, are $(u,v)$ and so we must have $R=ST=TS$. Moreover, for a
given isometry $u$ and $v$ are unique up to an overall sign. This
implies any $SO(2,2)$ matrix $R$ can be written as the product of two
commuting $SO(2,2)$ factors $S$ and $T$. However, the isometries
$(u^p,v^q)$ and $(u^{1-p},v^{1-q})$, whose product is $(u,v)$, also
commute, and so the split is not unique.

We define the 6 one-parameter subgroups $R_{[\mu\nu]}(\alpha)$ of
$SO(2,2)$, with $\mu>\nu$, $\mu,\nu=3,0,1,2$, as the group of
rotations or boosts in the $(x_\mu,x_\nu)$-plane. The subgroups
$R_{\mu\nu}$ and $R_{\kappa\lambda}$ commute if and only if
$\mu,\nu,\kappa,\lambda$ are all distinct, and so the 6 one-parameter
subgroups form three pairs of one-parameter commuting subgroups. We
can trivially combine these pairs into the 3 two-parameter subgroups
\begin{eqnarray}
R_{[30][12]}(\alpha,\beta)&:=&R_{[30]}(\alpha)\,R_{[12]}(\beta), \\
R_{[31][02]}(\alpha,\beta)&:=&R_{[31]}(\alpha)\,R_{[02]}(\beta), \\
R_{[32][01]}(\alpha,\beta)&:=&R_{[32]}(\alpha)\,R_{[01]}(\beta).
\end{eqnarray}
Less obviously, we can restrict these two obtain six more pairs of
two commuting one-parameter subgroups, namely
\begin{eqnarray}
\left[R_{[30][12]}(\alpha,\alpha),R_{[31][02]}(\beta,\beta)\right]&=&0, \\
\left[R_{[30][12]}(-\alpha,\alpha),R_{[31][02]}(-\beta,\beta)\right]&=&0, \\
\left[R_{[30][12]}(-\alpha,\alpha),R_{[32][01]}(\beta,\beta)\right]&=&0,\\
\left[R_{[30][12]}(\alpha,\alpha),R_{[32][01]}(-\beta,\beta)\right]&=&0,
\label{mycommutator}\\
\left[R_{[32][01]}(\alpha,\alpha),R_{[31][02]}(\beta,\beta)\right]&=&0,\\
\left[R_{[32][01]}(-\alpha,\alpha),R_{[31][02]}(-\beta,\beta)\right]&=&0.
\end{eqnarray}

 
\section{$SO(2,2)$ derivation of the overspinning particle
  cut-and-paste coordinates}
\label{appendix:SO22overspinning}


We now use the formulas of Appendix~\ref{appendix:SO22SL2R} to derive a
parameterisation of (\ref{hyperboloid}) for the overspinning case
$|J|>|M|$. In the context of this paper we need it to prove that
(\ref{overspinningplus},\ref{overspinningminus}) really are the
generators for $J>|M|$ and $J<|M|$, respectively.

The $SO(2,2)$ equivalent of the black hole generators
(\ref{BTZBHgenerators}) is
\begin{equation}
R_{\rm bh}:=R_{[32][01]}(2\pi s_-,-2\pi s_+).
\end{equation}
The two commuting factors corresponding to $(u,I)$ and $(I,v)$ are
\begin{eqnarray}
T_{\rm bh}&:=&R_{[32][01]}(-\pi\lambda_{+-},-\pi\lambda_{+-}), \\
S_{\rm bh}&:=&R_{[32][01]}(\pi\lambda_{++},-\pi\lambda_{++}).
\end{eqnarray}
The three parameterisations $X({\hat t},\chi,{\hat\phi})$ of
(\ref{hyperboloid}) for black holes can be written as
\begin{subequations}
\begin{eqnarray}
X_{\rm outer}&=&R_{[32][01]}({-\hat
  t,-\hat\phi})\,R_{[02]}(-\chi)\,(0,-1,0,0), \nonumber \\ \\
X_{\rm middle}&=&R_{[32][01]}({-\hat t,-\hat\phi})\,R_{[30]}(-\chi)\,(-1,0,0,0), \nonumber \\ \\
X_{\rm inner}&=&R_{[32][01]}({-\hat
  t,-\hat\phi})\,,R_{[31]}(-\chi)\,(-1,0,0,0). \nonumber \\ 
\end{eqnarray}
\end{subequations}

This notation makes it easy to verify that all three parameterisations
$X$ obey
\begin{equation}
R_{\rm bh}X({\hat t},\chi,{\hat\phi})=X({\hat t}-2\pi s_-,\chi,{\hat\phi}+2\pi s_+),
\end{equation}
using that $R_{32}$ and $R_{01}$ commute, and the defining property of
1-parameter subgroups that $R(\alpha)R(\beta)=R(\alpha+\beta)$.

Similarly, the equivalent of the point particle generators
(\ref{BTZparticlegenerators}) is
\begin{equation}
R_{\rm pp}:=R_{[30][12]}(-2\pi a_-,-2\pi a_+),
\end{equation}
with commuting factors 
\begin{eqnarray}
T_{\rm pp}&:=&R_{[30][12]}(-\pi\lambda  _{-+},-\pi\lambda_{-+}), \\
S_{\rm pp}&:=&R_{[30][12]}(\pi\lambda_{--},-\pi\lambda_{--}),
\end{eqnarray}
and the parameterisation is
\begin{equation}
X_{\rm pp}=R_{[30][12]}(-{\hat t},-{\hat\phi})\,R_{31}(\chi)\,(1,0,0,0), 
\end{equation}
and the pair obeys
\begin{equation}
R_{\rm pp}X_{\rm pp}({\hat t},\chi,{\hat\phi})=X({\hat t}+2\pi a_-,\chi,{\hat\phi}+2\pi a_+).
\end{equation}

Based on these examples, we now see what to do in the overspinning particle
case. We only deal with the case $J>|M|$, as the case $J<-|M|$ is
similar. The $SO(2,2)$ generator is
\begin{equation}
 R_{J>|M|}:=T_\text{pp}S_\text{bh}.
\end{equation}
The crucial observation is that the $T_\text{pp}$ and $S_\text{bh}$
still commute because of (\ref{mycommutator}). This allows us to write
\begin{eqnarray}
\label{Xoverspinning}
X_{J>|M|}&=&R_{[32][01]}(\hat\phi,-\hat\phi)
R_{[30][12]}(-\hat t,-\hat t) \nonumber \\
&&\times R_{02}(\chi)\,(0,1,0,0),
\end{eqnarray}
and to verify that $R_{J>|M|}X_{J>|M|}\sim X_{J>|M|}$ is equivalent to
(\ref{overspinningdefectangleandtimeshiftplus}). The
parameterisation (\ref{Xoverspinning}) is
written out in full in the last column of Table~\ref{table:BTZ}.
Instead of the last factor in (\ref{Xoverspinning}) we could have used
$R_{31}(-\chi)\,(1,0,0,0)$ to obtain the same induced metric.

 
\section{The limit $\Lambda=0$}
\label{appendix:Lambda0}


To take the limit $\Lambda\to 0$, or equivalently $\ell\to \infty$, we use
dimensional analysis to reinstate $\ell$ in the relevant formulas,
using that $J$ and our radial and time coordinates have dimension
length, while $M$ and angles are dimensionless. 

The metric of a spinning point particle solution of (\ref{EE}) with
$\Lambda=0$, with mass $m$ and spin (angular momentum) $j$ at rest at
the origin of the coordinate system can be written as the flat metric
\begin{equation}
\label{ppmetricDjtH}
ds^2=-d{\hat t}^2+d\chi^2+\chi^2\,d{\hat\phi}^2
\end{equation}
but with the non-trivial identifications
\begin{equation}
\label{spinningLambda0identification}
({\hat t},\chi,{\hat\phi})\sim\left({\hat t}+\Delta\hat t_{\rm
    pp},\chi,{\hat\phi}+(2\pi-2\nu)\right),
\end{equation}
where the defect angle $2\nu$ and time jump $\Delta\hat t_{\rm p}$ are
given by
\begin{eqnarray}
\label{nuppDJtH}
\nu&=&4\pi m, \\
\label{DeltathatDJtH}
\Delta\hat t_{\rm pp}&=&(1-4m)8\pi j,
\end{eqnarray}
in units $c=G=1$, compare Eqs.~(2.8b-c) and (4.21-22) of
\cite{DeserJackiwTHooft84} and Eq.~(1) of
\cite{DeserJackiwTHooft92}. Here $m$ and $j$ are defined as area
integrals over a distributional stress-energy tensor $T_{ab}$ in
Eq.~(\ref{EE}), following \cite{DeserJackiwTHooft84}. In the
limit $\ell\to\infty$, $\nu$ defined in (\ref{nuppdef}) and $\nu_\pm$
defined in (\ref{nupmdef}) all become become equal to $\nu$ defined in
(\ref{nuppDJtH}).

On small spacetime scales the cosmological constant becomes
irrelevant, and so has no influence on the conical singularity
itself. Indeed, in the limit $\ell\to\infty$, the
point particle metric (\ref{adS3metric}) in cut-and-paste coordinates
becomes (\ref{ppmetricDjtH}). 

Using $a_+\to\sqrt{-M}$ and $a_-\ell\to J/(2\sqrt{-M})$ (and hence
$a_-\to 0$ and $a_-/\ell\to 0$) as $\ell\to\infty$ we find that in
this limit the definitions (\ref{hatdefspp}) become
\begin{eqnarray}
\label{chitildedeflimit}
{\chi} &=&{1\over\sqrt{-M}}\sqrt{r^2+{J^2\over 4(-M)}}, \\
\
{\hat t} &=& \sqrt{-M}t +{J\over 2\sqrt{-M}} \phi, \\
{\hat\phi} &=& \sqrt{-M}\phi.
\end{eqnarray}
The resulting metric is again the BTZ metric (\ref{BTZmetric}), with the only
difference that now
\begin{equation}
\label{fLambda0}
f=-M+{J^2\over 4r^2}.
\end{equation}
This is agrees with taking the limit $\ell\to\infty$ directly in
(\ref{BTZmetric},\ref{fbetadef}) while keeping $M$ and $J$ fixed. 

We note that $f=0$ corresponds to $\chi=0$, whereas $r=0$ corresponds
to $\chi=|J|/(-2M)$. From the argument given above in
Sec.~\ref{section:CTCregion}, we see that by restricting to $r>0$ we
eliminate CTCs.

This important fact is not explicitly noted in
\cite{DeserJackiwTHooft84}, perhaps because the significance of
$\chi=|J|/(-2M)$ is less obvious in the cut-and-paste metric
(\ref{ppmetricDjtH}) than that of $r=0$ in the BTZ-like metric
(\ref{BTZmetric}) with (\ref{fLambda0}). Conversely, in
\cite{MiskovicZanelli2009} the point-particle BTZ metric is
interpreted as having a brane source at $r=0$, rather than a point
particle at $\chi=0$, and the possible extension beyond $r=0$ is not
explicitly noted.

The identification $(t,r,\phi)\sim(t,r,\phi+2\pi)$ corresponds
\begin{equation}
({\hat t},\chi,{\hat\phi})\sim\left({\hat t}+{J\over 2\sqrt{-M}}2\pi,\chi,{\hat\phi}+\sqrt{-M}2\pi\right),
\end{equation}
and so in the limit $\Lambda\to 0$ the point particle defect angle and
time jump are related to the BTZ quantities as
\begin{eqnarray}
\label{nuppdeflimit}
\nu&=&\pi\left(1-\sqrt{-M}\right), \\
\label{Deltathatdeflimit}
\Delta\hat t_{\rm pp}&=&{\pi J\over \sqrt{-M}}.
\end{eqnarray}
These expressions are equal to the leading orders in $\Lambda=-\ell^{-2}$ in the
corresponding expressions (\ref{nuppdef}) and
(\ref{Deltathatdef}). However, in interpreting them one should keep in
mind that $\nu$ and $\Delta\hat t_{\rm pp}$ can be measured in an
arbitrarily small neighbourhood of the particle world line, and so
should be interpreted as fundamental properties of the particle,
independently of $\Lambda$, while $M$ and $J$
should be considered as depending on both the particle and $\Lambda$.

As a byproduct of our calculations for $\Lambda<0$, we can also obtain
the expressions for the total mass and angular momentum of the effective
final particle in the case $\Lambda=0$, when black holes cannot
form. For this purpose, we simply note that the only dimensionful
parameter of the initial data is $D$, which has units of length and so
is measured in units of $\ell$. Among the parameters of the final
state, the same holds for the time jump $\Delta\hat t$.  Equating
(\ref{productmassivespinning}) with (\ref{pptracesbis}), taking the sum and
difference of the equations with the two signs, and expanding to
leading order in $1/\ell$, we find
\begin{eqnarray}
\label{Lambda0nu}
\cos\nu_\text{tot}&=&\cos\nu_1\cos\nu_2 -\sin\nu_1\sin\nu_2\cosh Z,
\nonumber \\ \\
\Delta\hat t_\text{tot}\sin\nu_\text{tot}&=&
\Delta\hat t_1(\sin\nu_1\cos\nu_2+\cos\nu_1\sin\nu_2\cosh Z)
\nonumber \\
&&+(1\leftrightarrow2)+2D\sin\nu_1\sin\nu_2\sinh Z. \nonumber \\
\end{eqnarray}
Note that the first equation is valid also for spinning point
particles, even though the time jumps representing the spin do not appear.

Eq.~(\ref{Lambda0nu}) is the the same as Eq.~(5.12) of
\cite{DeserJackiwTHooft84}, whereas the angular momentum $j$ of the
effective point particle was computed there only in a Newtonian
approximation. To check agreement with that result, we expand in
$\nu_i\ll 1$ and $Z\ll 1$. Noting that $Z\simeq {v_1+v_2}$ for small
$Z$, we find
\begin{eqnarray}
\nu&\simeq&\nu_1+\nu_2, \\
\Delta\hat t_{\rm pp}&\simeq&2{\nu_1\nu_2\over\nu_1+\nu_2}D\,(v_1+v_2).
\end{eqnarray}
For $\nu_1=\nu_2$ and hence $v_1=v_2$, this last equation reduces to
$\Delta\hat t_{\rm pp}\simeq 2\nu_1Dv_1$, which agrees with Eq.~(5.20)
of \cite{DeserJackiwTHooft84} if we also approximate the right-hand
side of (\ref{DeltathatDJtH}) by $8j$. (Note that in units $c=G=1$,
$J$ of \cite{DeserJackiwTHooft84} is our $j$.)

We note that the authors of \cite{DeserJackiwTHooft84} impose as the
rest frame condition that the effective particle is at rest in one of
the two product orders (but therefore not the other), whereas our rest
frame condition is that the effective particle moves with equal and
opposite velocities in the two product orders. We believe our
condition, which is based on the clear physical argument set out in
Sec.~\ref{section:restframecm}, is better motivated.

 
\section{A geometric construction of an effective particle with $\Lambda=0$}
\label{appendix:effectiveparticle}

\begin{figure}
\includegraphics[scale=0.65, angle=0]{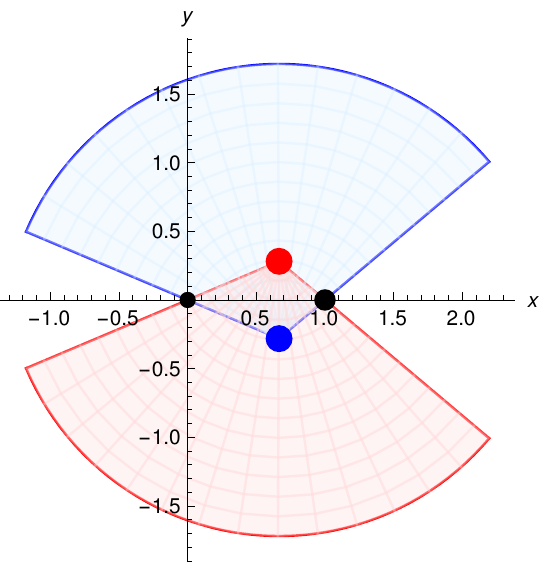} 
\caption{Construction of the spatial geometry with $\Lambda=0$ and two
  point particles with masses (half of defect angles) $\nu_1$ and
  $\nu_2$ at positions $(0,0)$ and $(0,d)$, represented by the two
  smaller (black) dots. The area of each dot is proportional to the
  mass (defect angle) of the particle it represents. For plotting we
  have chosen the specific values $\nu_1=0.4$, $\nu_2=0.7$, and
  $d=1.0$.}
\label{figure:fig1}
\end{figure}

Fig.~\ref{figure:fig1} illustrates a time slice with $\Lambda=0$ and
two point particles with masses (half of defect angles) $\nu_1$ and
$\nu_2$ at rest at positions $(0,0)$ and $(0,d)$.  The following
statements can be verified by inspection of this figure:

(1) A time slice is obtained by gluing
together the entire shaded region (that is, counting the overlap only
once) together along the two wedges. The resulting geometry, when
embedded like a piece of paper into 3-dimensional Euclidean space,
looks like a coffee filter paper. Let us call this manifold~1.

(2) By contrast, the effective geometry seen from a distance is
obtained by moving the upper (blue) and lower (red) shaded patches
apart until their vertices (represented by the two larger dots, with
the corresponding colors) coincide before gluing them together. The
total defect angle at the new vertex (where the two larger dots are
now identified) is $2\nu_1+2\nu_2$. Note that the doubly-shaded region
of space has to be duplicated for this construction. Let us call the
result manifold~2.

(3) The new vertex has a unique position on manifold~2
(which is larger), but not on manifold~1.  If on each of
the two patches on manifold 1 we keep the Cartesian coordinate system
$(x,y)$ defined by the horizontal and vertical axes before the patches
are moved then the location of the effective particle is
\begin{equation}
(x_0,y_0):=d{\sin\nu_2\over \sin(\nu_1+\nu_2)}(\cos\nu_1,\sin\nu_1)
\end{equation}
when seen from the lower (red) patch, but $(x_0,-y_0)$ when seen from
the upper (blue) patch.

(4) Seen from the upper (blue) patch, going anticlockwise around the
effective particle is equivalent to going anticlockwise around
particle~1 then particle~2. Seen from the lower (red) patch, this
order is reversed. Hence the two product orders correspond to
different coordinate choices on the two-particle spacetime.

By making $d$ slowly time-dependent, we can also infer the geometry of
two moving particles in the limit of non-relativistic motion: If the
particles approach each other horizontally at a combined speed $2v$,
then the two fictitious locations of the effective particle approach
each other vertically, each at speed of
$v_\text{tot}=2v\sin\nu_1\nu_2/\sin(\nu_1+\nu_2)$. For $\nu_1=\nu_2=\nu$ and
$v_1=v_2=v$, we have $v_\text{tot}=v\tan\nu$. This agrees with the
non-relativistic limit $|v|\ll 1$ of (\ref{zf}), and
$\nu_\text{tot}=\nu_1+\nu_2$ agrees with (\ref{nuf}). (At higher velocity,
special-relativistic velocity addition and mass increase would have to
be taken into account).

\section{The head-on collision of two non-spinning massive particles}
\label{appendix:Steif}


Steif \cite{Steif96} has classified all time-symmetric initial data
for the two-body problem. The two bodies can then be nonspinning black
holes or nonspinning massive point particles.

For the case of two point particles, we have the 3-dimensional subcase
of our 6-dimensional space of initial data where two non-spinning
massive particles collide head-on. This requires that the particle
spins and the impact parameter $D$ all vanish. In our
units where $\Lambda=-1$ and $G=c=1$, our
parameter $M$ is $8M$ of Steif, our $\nu_1$ and $\nu_2$ are $4\pi m$
and $4\pi\tilde m$ of Steif, and our $Z$ is $d$ of Steif. Recall
that $Z$ is both the rapidity at closest approach (here, at impact),
and the distance at largest separation (here, at the moment of time
symmetry). Rather than the full spacetime, Steif constructs the
initial data at the moment of time symmetry, where the extrinsic
curvature vanishes. The spatial geometry at that moment is constructed
by isometric identifications of the Poincar\'e disk, corresponding to
our identifications of the entire spacetime. Note that for easier
comparison with Steif's work we have adapted a convention throughout
this paper where $u=v$ if and only if the spacetime admits a moment of
time symmetry, and this single element of $SL(2,\R)$ then
parameterises an isometry of ${\mathbb H}^2$.

Our results agree with Steif's with the following clarifications: his
``open, no horizon'' can denote both our virtual effective particle
outcome, or our black hole outcome. In the latter case, a black hole
forms eventually as the particles fall toward each other from a large
separation, but there is no apparent horizon at the moment of time
symmetry (moment of largest separation). ``Closed space'' corresponds
to our real effective particle outcome. In other words, this is really
a three-particle closed universe that recollapses. ``Black hole''
corresponds to our black hole outcome {\em and} where an apparent
horizon is already present at the moment of time symmetry. Our black
hole criterion $Z>C$ corresponds to $\cosh d>f_c$ of Steif. Our and
Steif's results are shown in Fig.~\ref{figure:Steif2particleplot}. For
completeness, we have included in this figure, without derivation,
Steif's criterion for the existence of an AH at the moment of time
symmetry. This is $d<\tan\nu_1/(-\tan\nu_2)$ for $0\le\nu_1\le\pi/2$
and $\pi/2<\nu_2<\pi$, and equivalently for $\nu_1$ and $\nu_2$
interchanged. The derivation, not given in \cite{Steif96}, will be
given elsewhere \cite{Jormatwoparticles}.

When the two original particles are equivalent to an effective
particle and collide head-on but a black hole does not form, one can
take them to collide elastically, bouncing back from the collision
event, or to collide inelastically, continuing as a single particle with
mass $\nu_\text{tot}\pi$.

In the elastic collision, one basically wants to glue one spacetime
picture, including the wedges, to its mirror image under a time
reflection through $t=0$. This is possible only if the restrictions of
the two sides of each wedge to $t=0$ are mapped to each other, and
this in turn requires the wedges to be centred on the plane through
the two particle trajectories, and facing away from each other.

For the inelastic collision, one could arrange one side of each
excision wedge to touch at the moment of collision, and for the other
two sides to become, at that moment, the two sides of the excision
wedge of the new particle.


\end{appendix}



\end{document}